# Rare earth permanent magnets prepared by hot deformation process[*]

Chen Ren-Jie(陈仁杰)[†], Wang Ze-Xuan(王泽轩), Tang Xu(唐旭), Yin Wen-Zong(尹文宗), Jin Chao-Xiang(靳朝相), Ju Jin-Yun(剧锦云), Yan A-Ru(闫阿儒)[‡]

CAS Key Laboratory of Magnetic Materials and Devices, Ningbo Institute of Material Technology and Engineering, Chinese Academy of Sciences, Ningbo 315201, People's Republic of China

**Abstract**

Hot deformation process is one of the primary methods to produce anisotropic rare earth permanent magnets. Firstly, rapidly quenched powder flakes with nanocrystal structure are condensed into the full dense isotropic precursors by hot pressing process. And then, the prepared isotropic precursors are hot deformed to produce high-anisotropy uniaxial bulk rare earth permanent magnets, in which the highly textured structure is obtained in the hot plastic deformation process. The obtained hot-deformed magnets possess many advantages, such as near net-shape, outstanding corrosion resistance and ultrafine-grain structure. The noteworthy effects of preparation parameters employed in hot-pressing and deformation processes on the magnetic properties and microstructures characterizations are systemically summarized in this academic monograph. As a near net-shape technique, hot deformation process has noteworthy advantages in producing irregular shape magnets, especially for radially oriented ring-shape magnets with high length-diameter ratio or thin wall. The difficulties in producing crack-free, homogeneous and non-decentered ring-shaped magnets are basically resolved through mold design, adjustment of deformation parameters and application of theoretical simulation. Considering the characteristics of hot-deformed magnets, such as the grain shapes and sizes, anisotropic distribution of intergranular phases, etc., there is practical significance to study and improve the mechanical, electric properties and thermal stability to enlarge the applicable area of hot-deformed magnets or ring-shaped magnets.

[*] Project supported by the National Key Research and Development Program of China (Grant No. 2016YFB0700902), the National Natural Science Foundation of China (Grants No. 51671207, 51601207, 51501213)
[†] Corresponding author. E-mail: chenrj@nimte.ac.cn
[‡] Corresponding author. E-mail: aruyan@nimte.ac.cn

# Section 1 Introduction

In the mid-1980s, R. W. Lee [1] originally reported a method to prepare an anisotropic Nd-Fe-B magnet with ultrafine grain size by hot pressing and die-upsetting the rapidly quenched Nd-Fe-B alloy at a high temperature. Compared to the traditional sintering method, the die-upsetting process can realize a strong magnetic anisotropy without applying an external magnetic field, which makes it appropriate to produce the ring magnets with radial orientation. The formation of texture in this kind of Nd-Fe-B magnet, i.e. hot deformation (HD) magnet, is attributed to the anisotropy of the elastic modulus of $Nd_2Fe_{14}B$ grains. As shown in Fig. 1[2], the elastic modulus along c axis ($E_{//}$) is much smaller than that perpendicular to c axis ($E_\perp$). And the difference increases rapidly when the temperature is higher than 500 °C. Therefore, the c-axes of $Nd_2Fe_{14}B$ grains are inclined to parallel to the pressure direction at high temperature. In fact, the detailed mechanism of alignment or texture formation is more complicated in a real system. It is generally acknowledged that the high anisotropy is attributed to both the rotation and the preferred growth of $Nd_2Fe_{14}B$ grains under the pressure at high temperature. The grains, whose c-axes have smaller angle with the press direction, are easier to "devour" the misoriented grains via the "precipitation-growth" or "diffusion slip" processes, because the strain energy of the misoriented grains is much higher than that of well-aligned grains [3, 4]. The preferred growth and rotation of grains may be simultaneous, or more likely, they are dominating in different steps in the texture formation process. As a special example, the radial alignment for back extruded ring magnet is obtained by the platelet-shaped grains rotation or alignment in the extrusion process and the platelet-shaped grains are already formed before extrusion [5]. Obviously, rare-earth-rich intergranular phase (liquid phase at high temperature), i.e., rare earth content greater than stoichiometric ratio of $Nd_2Fe_{14}B$, is an essential factor to obtain the platelet-like shape grains and magnetic anisotropy of the whole magnet. Of course, the plastic deformation of the magnet in the die-upsetting process is necessary as well. Fig. 2 shows the demagnetization curves of over-stoichiometric $Nd_{13.34}Fe_{74.64}Co_{5.5}Ga_{0.42}B_{6.0}Nb_{0.1}$ and rare-earth-lean $Nd_{7.91}Pr_{2.68}Fe_{84.01}B_{5.39}$ (at. %) hot-pressing (HP) and HD magnets.

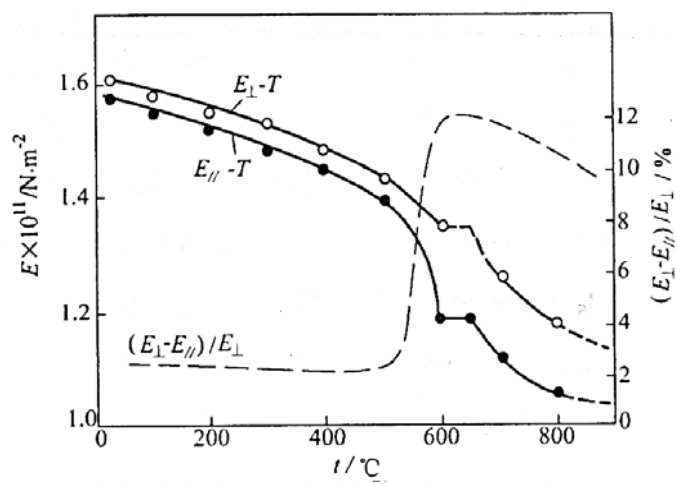

Fig. 1. Variations of elastic modulus of $Nd_2Fe_{14}B$ along ($E_{//}$) and perpendicular ($E_\perp$) to c-axis directions and their difference on temperature [2].

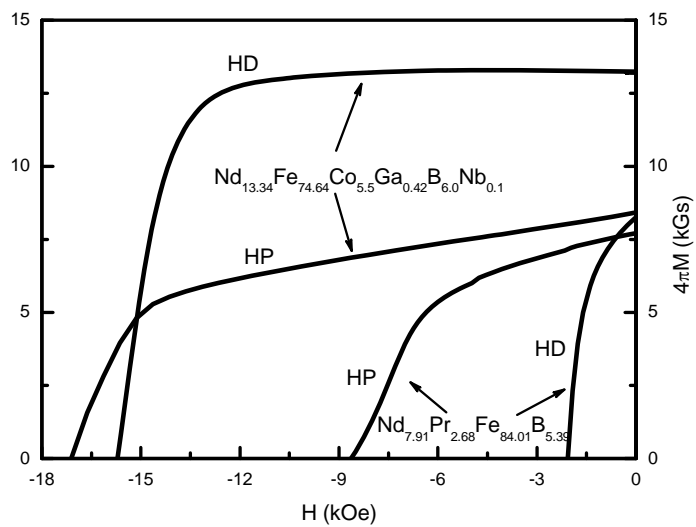

Fig. 2. Demagnetization curves of HP and HD of the over-stoichiometric and rare-earth-lean magnets.

There are three advantages of hot deformation process concluded by Croat *et al.* [6]: fewer manufacturing steps suitable for a continuous mass production; net shaping or near net shaping requiring little or no finish grinding; ideal process for producing special shape magnet such as arc or ring shape with radial orientation. Up to now, the backward extruded ring magnets with radial orientation have been used in electric power steering system and AC servo motor. The HD magnets with an ultrafine-grained structure are potential to obtain high coercivity ($H_{cj}$), which makes them regain much attention. Recently, the researches about hot deformed Nd-Fe-B magnets are mainly focused on the preparation technology and microstructure optimization to improve magnetic properties, especially coercivity. Besides, the stress-induced crystallographic orientation is another critical issue for hot-deformed Nd-Fe-B magnets. Besides, hot-pressing and hot deformation processes have prominent superiority on producing bulk nanocomposite permanent magnet which is once considered to be the candidate of the fourth generation rare earth permanent magnets. But the previous researches indicate that texture formation is very difficult by hot deformation process for nanocomposite magnets due to the rare-earth-lean composition. By infiltrating the eutectic alloys into the grain boundary of nanocomposite magnets, the abilities of deformation and texture formation are improvemed remarkably. The orientation mechanism of the diffusion-processed nanocomposite magnets is also discussed in recent years.

**Section 2 Effects of preparation parameters on HD magnets**
The preparation process including melt spinning, hot-pressing and hot deformation significantly influences the microstructure and magnetic properties of HD magnets. As the starting material, the powders crushed from the melt-spun ribbons consist of nanocrystal [7] or amorphous matrix [8]. Although melt-spun ribbons exhibit high oxidation resistance, the oxygen content of the HD magnet increases with decreasing the powders size. In order to produce hot deformed magnets with better performance, the particle size of commercial melt-spun powders is generally 200 - 350 μm [9]. Besides, the microstructure and magnetic properties of hot-deformed magnets have close relationship with hot pressing and hot deformation process. Hereby, the influences of preparation parameters on the microstructure and magnetic properties of magnets will be discussed in detail, such as hot pressing temperature, hot deformation temperature, deformation degree and strain rate.

2.1 HP temperature

Isotropic precursor is obtained by hot-pressing melt-spun powders. The density variation of the precursors prepared at different HP temperatures is shown in Fig. 3(a). The density monotonically increases with the increase of HP temperature before 670 °C and reaches nearly full density (~ 7.6 g/cm$^3$) with further increasing temperature. Fig. 3(b) illustrates the demagnetization curves of HP magnets prepared at different temperatures. When HP temperature increases from 630 to 710°C, the remanence $B_r$ remains at about 8 kG because of magnetic isotropy. The squareness of demagnetization curve for the sample prepared at 630 °C is distinctly inferior to those at 670 and 710 °C, which can be well explained by the density variation. The coercivity of samples prepared at 630 and 670 °C is ~ 18.5 kOe, slightly higher than the one prepared at 710 °C. The microstructure of HP precursor is sensitive to elevated temperature while the size of grains keeps at 30~80 nm [10]. The isotropic fully dense precursor with small grains is favourable to the magnetic properties [11]. Further elevated HP temperature results in the formation of coarse grains and consequent reduces of hot workability, seriously weakening c-axis texture in HD magnet. Li *et al* [12] reported that microstructure of HD magnets is closely related with the starting microstructure of their precursors. In addition, Liu *et al.* [13] pointed out that magnets prepared at elevated temperature maybe give rise to the extrusion of partial RE-rich phase, which reduces the coercivity because of the weakened decoupling between grains.

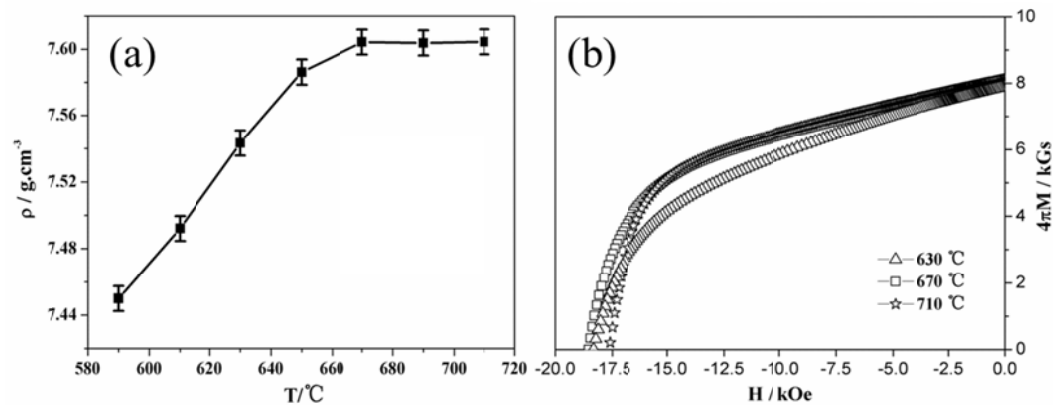

Fig. 3. (a) Dependence of the density of HP precursor on HP temperature ranging from 590 to 710 °C and (b) demagnetization curves of HP precursors prepared at the temperatures of 630, 670 and 710 °C.

## 2.2 HD temperature

The fully dense precursors are deformed into anisotropic magnets with good texture during HD process under suitable temperature and stress [9, 10]. The texture in HD Nd-Fe-B magnet is developed because of stress-induced preferential grain growth via dissolution-precipitation mechanism [14, 15]. Fig. 4 shows the microstructure of hot-deformed magnets prepared at different HD temperatures from 780 to 860 °C, observed by scanning electron microscope (SEM). For the magnet deformed at 780 °C, equiaxed grains and fine platelet-shaped grains with inferior alignment coexist, as shown in Fig. 4(a). It indicates that both the growth and orientation of $Nd_2Fe_{14}B$ grains are inhibited in HD process due to the relatively low temperature. When HD temperature ascends to 820 and 840 °C, well-aligned platelet-shaped grains with the thickness of 50 - 70 nm and length of 300 - 700 nm are formed (Fig. b and 4c). With HD temperature further increasing to 860 °C, the grain size obviously increases (Fig. d).

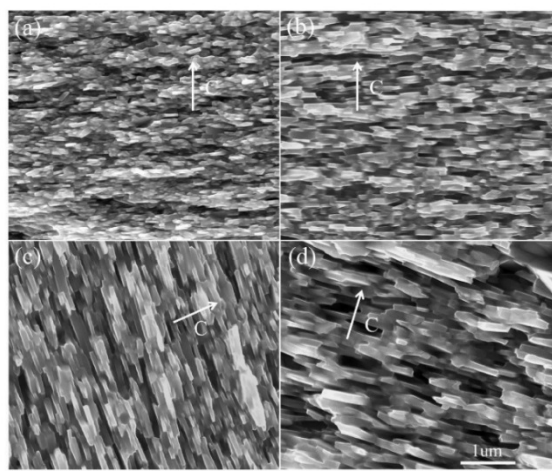

Fig. 4. SEM photos of HD magnets prepared at the HD temperature of (a) 780, (b) 820, (c) 840, and (d) 860°C. The white arrows represent the easy magnetization direction (c axis) that is perpendicular to the plane of platelet grains.

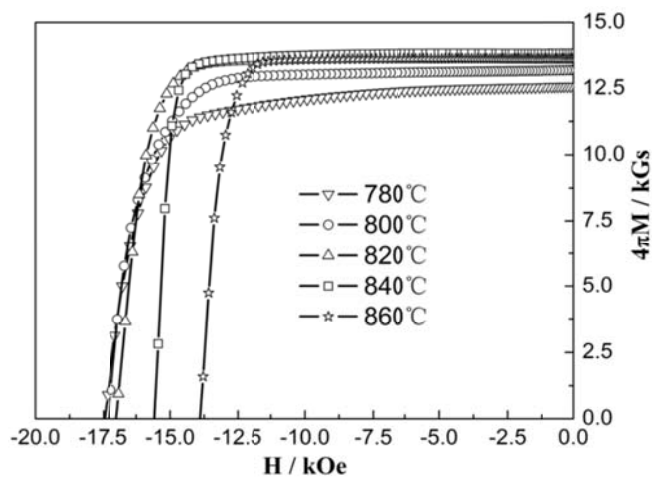

Fig. 5. Demagnetization curves of HD magnets prepared at different HD temperatures.

Demagnetization curves of HD magnets prepared at the HD temperatures of 780 - 860 °C are shown in Fig. . It can be seen that remanence increases with the HD temperature before 840 °C and then decreases with the HD temperature further increasing to 860°C. For the magnet deformed at 780 °C, the low remanence is due to the stronger deformation resistance and consequent lower orientation. When HD temperature increases to 820 - 840 °C, the platelet-shaped $Nd_2Fe_{14}B$ grains are well-aligned. This is consistent with the superior value of remanence. Coercivity decreases monotonically with the increase of HD temperature because of the grain growth. Based on the above analysis, the deformation temperature is optimized to be 820 - 840 °C. The remanence, coercivity and maximum energy product of the HD magnet deformed at 840 °C are 13.8 kG, 15.8 kOe and 46.5 MGOe, respectively. Lin *et al.* [16] reported that the abnormal grain growth and Nd-rich phase extrusion occur when HD temperature is higher than 850 °C, which leads to the reductions of density and remanence. Zhao *et al.* [17] reported that the effective anisotropy can be obtained at the HD temperature of 600 - 900 °C, and the suitable HD temperature is proved to be 800 °C in their work. It should be noted that the suitable HD temperature varies with different heating methods. The optimum HD temperature of 680 °C has been reported by Hu *et al.* [18] using spark plasma sintering as deformation method.

2.3 Deformation degree

Hot deformed magnets with the deformation degrees from 30% to 80% are produced and their magnetic properties are investigated (Fig. 6a). With the deformation degree raised from 30% to 70%, coercivity decreases from 17.3 kOe to 15.3 kOe while remanence increases from 10.6 kGs to 13.75 kGs. When the deformation degree further increases to 80%, coercivity decreases rapidly and remanence starts to decline. Maximum energy product presents similar variation to remanence and the maximum value is 45.8 MGOe at the deformation degree of 70%. The XRD patterns of the HD magnets with the deformation degree of 30, 70 and 80% are given in Fig. 6(b). For the 30%-deformation magnet, the intensity of (004), (006) and (008) peaks are relatively weak, indicating the inferior texture. When the deformation degree is up to 70 - 80%, the intensity of (00*l*) peaks becomes much stronger. The (006) peak shows higher intensity than (105) peak, which demonstrates a strong texture in the HD magnet. Because of the enhancement of texture, larger remanence and maximum energy product are obtained in the HD magnets with higher deformation degree.

Fig.7Fig.  shows the microstructure of hot deformed magnets with the deformation degree of 30, 70, and 80%. In the 30%-deformation magnet, the equiaxed grains and the platelet-shaped grains with poor alignment coexist due to insufficient deformation degree (Fig. 7a). In the 70%-deformation magnet, fine platelet-shaped grains with superior alignment are formed, as shown in Fig.  (7b). When deformation degree increases to 80%, platelet-shaped grains still exhibit fine alignment but grain size is up to 1 μm in length (Fig. c), resulting in the remarkable reduction of coercivity. Based on these experimental results, it is concluded that the best deformation degree

for the optimal magnetic properties is about 70%. However, deformation degree is optimized to be 75% in Zhao's research [17] and 80% in Liu's [19], respectively. In Zhao's work, higher deformation degree leads to non-uniform deformation and defects such as the microcracks between melt spun ribbons in magnet [17]. When well-aligned $Nd_2Fe_{14}B$ platelet is realized in HD magnet, it is difficult to further enhance remanence merely by raising deformation degree. In addition, optimal deformation degree of HD magnet commonly depended on the composition.

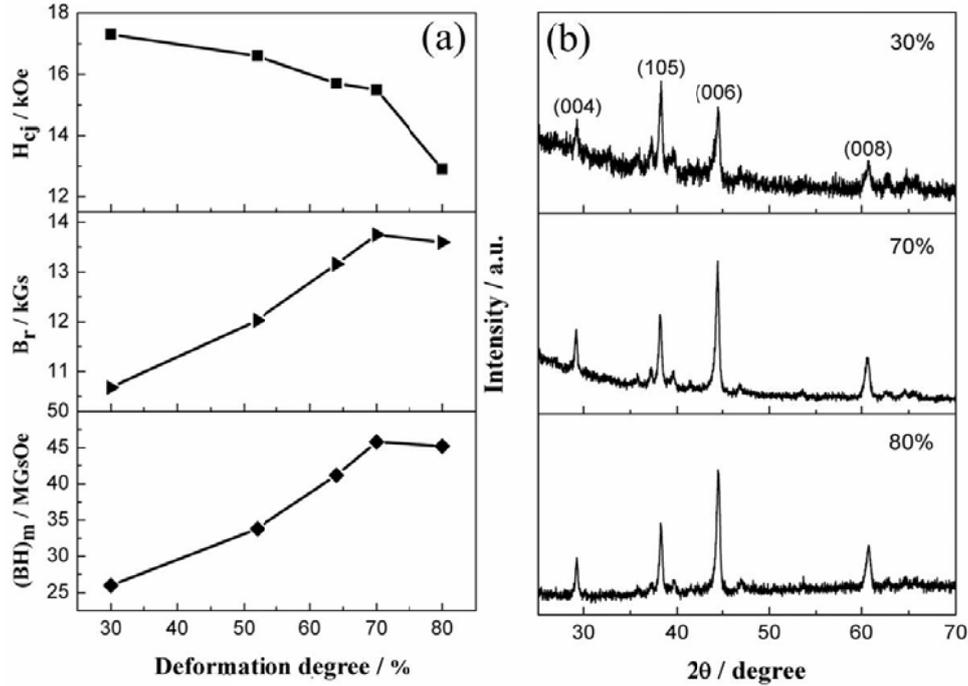

Fig. 6. (a) Magnetic properties of hot deformed magnets with the deformation degree of 30, 52, 64, 70 and 80%; (b) XRD patterns of the hot deformed magnets with the deformation degree of 30, 70 and 80%.

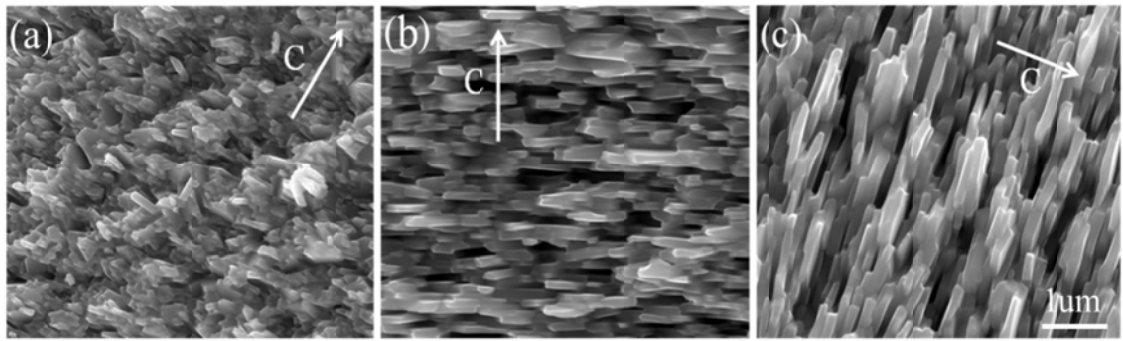

Fig. 7. Microstructure of hot deformed magnets with the deformation degree of (a) 30, (b) 70 and (c) 80%. White arrows represent the easy magnetization direction of c axis that is perpendicular to the plane of platelet grains.

2.4 Strain rate

The deformation speed can be evaluated with constant strain rate $\dot{\varphi} = d\varphi/dt$, where $\varphi = ln(h_0/h)$, $h_0$ is the starting height of HP precursor and $h$ is the ultimate height of hot deformed magnets [17, 20]. Strain rate affects the microstructure and magnetic properties of hot deformed magnets substantially. Fig. shows the dependence of magnetic properties of hot-deformed magnets on strain rate. Coercivity increases monotonically with the increment of strain rate. This is ascribed to the finer grains at higher deformation speed due to shorter time exposed to high temperature. Both remanence and maximum energy product ascend firstly and then descend quickly with the rise of strain rate. The best strain rate is proved to be 25 × $10^{-3}$/s. Higher strain rate leads to less deformation of grains and inferior texture. Zhao et al. [17] reported that the maximum energy product and remanence are obtained at the strain rate of $10^{-2}$ $s^{-1}$. This suggests that lower strain rate brings about longer deformation time and results in lower coercivity for grain coarsening. In fact, too slow deformation also causes the reduction of remanence because the non-aligned coarse grains prevent the formation of c-axis texture originating from fine grains.

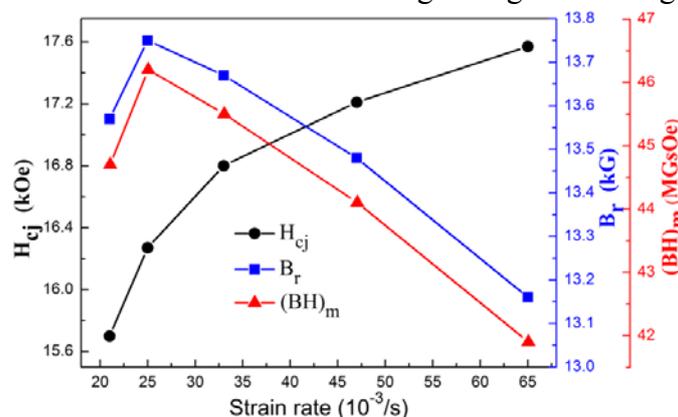

Fig. 8. Magnetic properties of hot deformed magnets prepared with different strain rates.

2.5 The inhomogeneity of microstructure in HD magnet

Coarse grain region and fine equiaxed grain region can be found in hot deformed magnets as shown in Fig. 9 (a) and (b). Both of them grow rapidly with the ascending of deformation temperature [16] and deformation degree [21]. The two kinds of grains are disadvantageous to magnetic properties. Especially, the coarse grains can lower remanence. The coarse grain regions contain higher Nd content and mainly form at the ribbon boundaries with quasi-periodic distribution. Li et al. [12] reported that the growth of coarse grains initiates in HP precursor at high hot compaction temperature and propagates during the subsequent hot deformation. Lai et al. [22] reported that coarse grain region is correlated to the incomplete contact interfaces of neighboring ribbons. Under high temperature and stress, Nd-rich phase is squeezed into the interspaces of neighbor ribbons and stress concentration exists in the contacted points. Because of the fluidity of Nd-rich phases and the release of interfacial energy, the non-aligned coarse grain regions form both in HP and HD process.

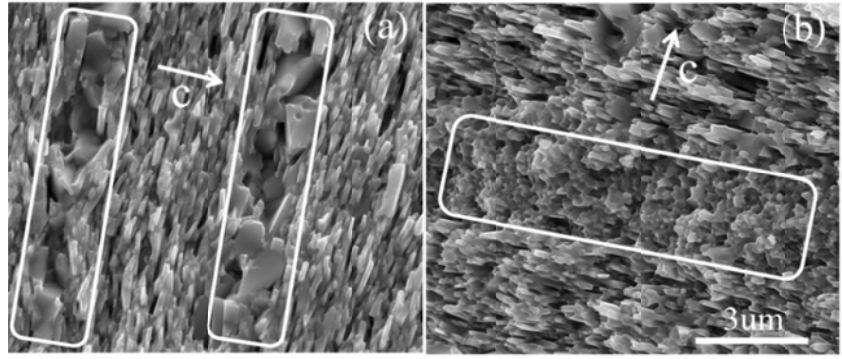

Fig. 9. SEM images of (a) coarse grain regions and (b) fine equiaxed grain regions.

Fig. 10 shows the further results on the quantitative variation of magnetic properties along axial direction over the HD magnets with different deformation degrees. The remanence of sample 2 and 3 is significantly higher than that of sample 1 and 4. The inhomogeneity of remanence indicates the inhomogeneity of deformation degree along axial direction in HD magnets. In addition, the inhomogeneity of microstructure also exists along the radial direction of magnet [23].

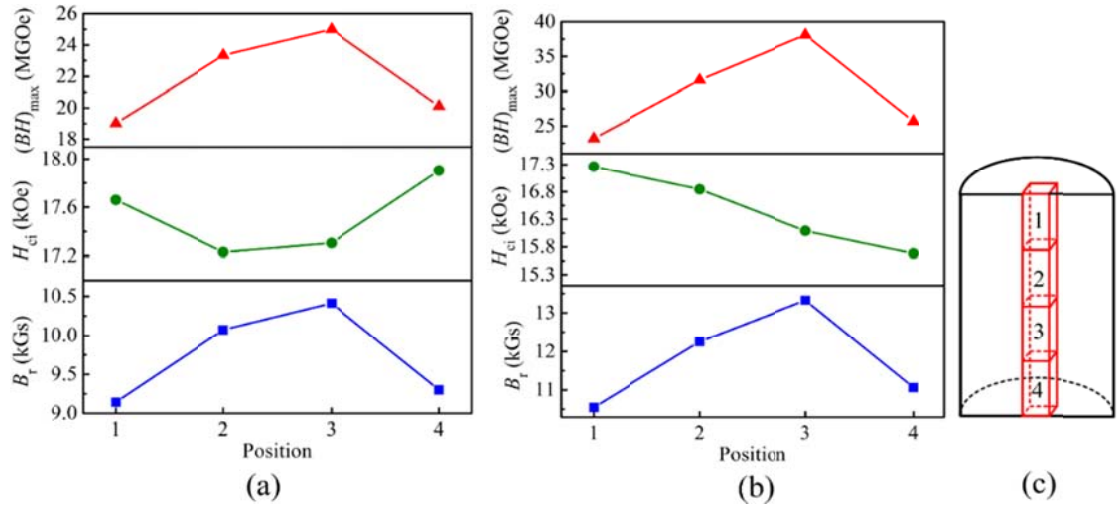

Fig. 10. The distribution of remanence ($B_r$), coercivity ($H_{cj}$) and maximum energy product ($(BH)_{max}$) along axial direction over the HD magnet with the deformation degree of (a) 37 and (b) 52% [23].

2.4 The influence of chemical composition

Hot deformed Nd-Fe-B magnets are multi-phase materials which consist of $Nd_2Fe_{14}B$ matrix phase, Nd-rich phase and B-rich compound. Magnetic properties, especially remanence and energy product, strongly depend on the grains alignment along c-axis direction and the content and distribution of Nd-rich phase. In the deformation process, $Nd_2Fe_{14}B$ grains are wetted by liquid Nd-rich phase, which promotes the formation of platelet-shaped grains by grain slipping and rotation [9, 24]. In addition, Nd-rich phase at grain boundaries enhances the decoupling effect of $Nd_2Fe_{14}B$ grains and facilitates the augment of coercivity. Therefore, the Nd content in hot deformed magnet is higher than that in $Nd_2Fe_{14}B$ tetragonal phase. Yi *et al.* [4]

investigated the influence of Nd content on the microstructure and magnetic properties of hot deformed magnets with nominal composition $Nd_{12.8+x}Fe_{75.28-x}Co_{5.5}Ga_{0.42}B_6$, as shown in Fig. 11 and Fig. 12. With increasing the Nd content, remanence and energy product increase until x = 0.54 and then declines (Fig. 11). Coercivity increases monotonically with x addition from 0 to 1. As shown in Fig. 12(a), evident equiaxed grains exist in the Nd-lean alloy (x=0). As shown in Fig. 12(b), the main grains are elongated and a good texture is obtained in Nd-rich magnet with x = 0.54. However, the parallel alignment of $Nd_2Fe_{14}B$ grains is strongly distorted and circular alignment is formed in the excessive Nd-rich alloy with x = 1.08, as shown in Fig. 12(c). Similarly, Liu [13] produced Nd-Fe-B magnets with the coercivity of 17.9 kOe by increasing the Nd content and reveals that pinning mechanism plays a dominant role in determining the coercivity. In Nd-Fe-B magnetic materials, Dy [25] and Tb [26] are the most commonly used addition to improve coercivity due to the superior magnetocrystalline anisotropy fields (Ha). Tang *et al.* [27] ascends the coercivity of Nd-Fe-B hot-deformed magnet from 15.0 to 22.7 kOe by addition of Dy-based melt-spun powders.

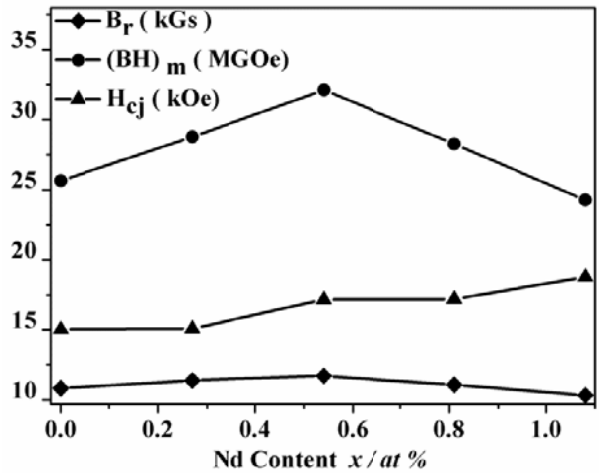

Fig. 11. Magnetic properties of the $Nd_{12.8+x}Fe_{75.28-x}Co_{5.5}Ga_{0.42}B_6$ magnets with different Nd content [4].

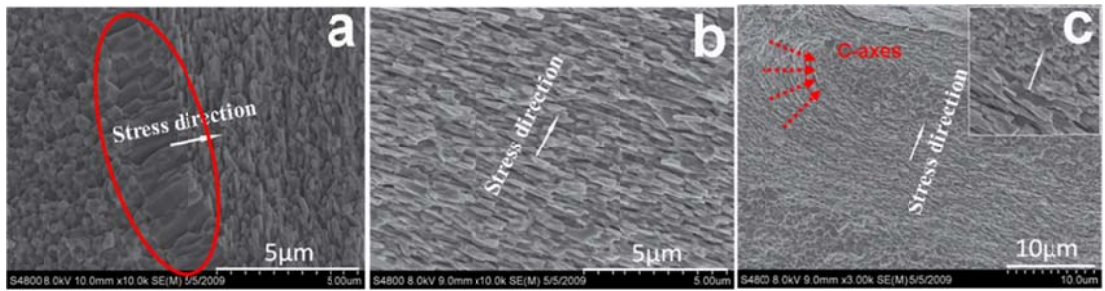

Fig. 12. Microstructure of $Nd_{12.8+x}Fe_{75.28-x}Co_{5.5}Ga_{0.42}B_6$ magnets with (a) x=0, (b) x=0.54 and (c) x=0.18 [4].

It is reported that coercivity and maximum energy product are enhanced significantly by Cu powder addition using dual-alloys technology [14]. Microstructure analyses indicate that $NdCu_2$ phase is formed in grain boundaries, which obviously improves the texture of Cu-doped magnet due to the lower melting point of intergranular phase and leads to the enhancement of magnetic properties. Ma's research [28] indicates that Nb addition leads to remarkable increase of both remanence and coercivity because of the improvement of c-axis texture and refinement of microstructure. Microstructure analysis reveals that Nb atoms enrich at grain boundaries and form NbFeB compound. Li et al. [29] researched the influence of Al, $MoS_2$, Zn, etc. on the magnetic properties, and found that $MoS_2$ and Al are effective in increasing remanence and energy product and Zn can increase coercivity. Besides, Co, Ga, Zr, V, Mo, etc. are utilized for HD magnets to tune microstructure or improve magnetic properties [30-32].

In fact, $SmCo_5$ permanent magnet also can be fabricated by hot pressing and subsequent hot deformation process. In recent years, Yue [33, 34] reported hot-deformed anisotropic $SmCo_5$ magnets with a high coercivity over 5 T. The raw powder was prepared with ball milling technique. The result reveals that deformation level up to 90% leads to the formation of well-aligned platelet grains. The images on microstructure are shown in Fig.13 [33]. As observed in Fig. 13(a), the size of equiaxial grain is about 20 nm in HP sample. After deformation, the grains grow up and elongate perpendicular to press direction (arrows in Fig. 13(b) and (c)). However, partial equiaxial grains still exist in 70%-deformation magnet due to the inadequate deformation degree. When deformation degree increases to 90%, hot deformed $SmCo_5$ magnet mainly consists of aligned platelet $SmCo_5$ grains. Therefore, deformation degree plays a key role in inducing the grain alignment and the formation of platelet shape.

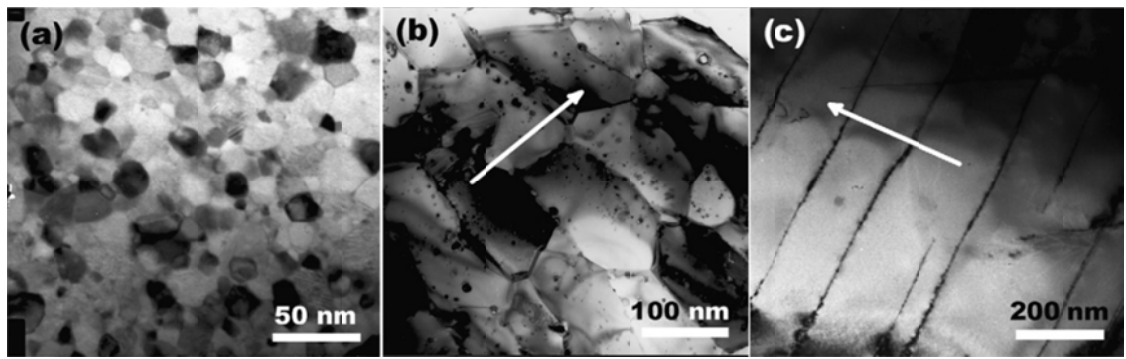

Fig. 13. TEM images of (a) hot pressed, (b) 70%-deformation and (c) 90%-deformation $SmCo_5$ magnets [33].

**Section 3 Coercivity enhancement for HD Nd-Fe-B magnets**

Coercivity is an intrinsic property which plays a key role in the commercial applications of Nd-Fe-B magnets. However, the maximum experimental value of coercivity is only 20 - 30% of the theoretical value which is deduced according to S-W model [35]. The enhancement of coercivity has been the primary challenge for researches on Nd-Fe-B magnets. Coercivity of Nd-Fe-B magnets is sensitive to microstructure such as defects on the surface of grains, ferromagnetic grain boundary phases and effective demagnetization factors. In order to find appropriate method to raise coercivity, researchers have taken much effort to clarify the coercivity mechanism of Nd-Fe-B magnets.

3.1 Coercivity mechanism of HD Nd-Fe-B magnets

Generally, there are two dominant coercivity mechanisms of Nd-Fe-B magnets: nucleation and domain wall pinning mechanisms. However, there is no final decision about the controversy between the two mechanisms to conclude which one is indeed suitable. In hot-deformed Nd-Fe-B magnets, a strong domain wall pinning is considered to be the primary coercivity mechanism [36, 37]. Theoretically, the strength of domain wall pinning force at grain boundaries determines the coercivity values. The initial magnetization curves are often adopted to discuss the magnetization behavior and coercivity mechanism. An S-shaped profile of the initial magnetization which indicates the existence of complicity in coercivity mechanism is obtained in HD magnets.

The first step shows high reversal susceptibility and relatively easy magnetization. It indicates that reversal domain wall motion dominate the magnetization behavior. In the second step, domain wall pinning at grain boundaries is dominant. The two steps S-shaped magnetization curve shows an inhomogeneous domain wall pinning mechanism in HD Nd-Fe-B magnets. In addition, Volkov *et al.* [37] investigated the magnetic domain walls structures in hot-deformed Nd-Fe-B magnets and discussed the domain wall pinning at grain boundaries and inhomogeneous microstructures in detail. The inhomogeneous microstructures which are grain boundary phases and different types of defects on grain surfaces would result in the inhomogeneous domain wall pinning. An inhomogeneous domain wall pinning mechanism is relatively appropriate to describe the dominant coercivity mechanism in HD Nd-Fe-B magnets. However, a nucleated micromagnetic model ($H_c(T)= \alpha H_A(T)-N_{eff}M_s(T)$) is always applied to analyze the coercivity mechanism in HD Nd-Fe-B magnets. A linear relationship of $H_c(T)/M_s(T)$ vs. $H_A(T)/M_s(T)$ is obtained so that the coercivity mechanism of hot-deformed magnets seems to be close to the nucleation mechanism as well.

The enhancement of coercivity has been the primary challenges for researchers. In Nd-Fe-B magnets, it has reported that the coercivity will increase along with the decrease of grain sizes. According to the S-W model, the maximum coercivity could be obtained as soon as the grain sizes are single domain sizes and coherent rotation of the magnetic moments is perfectly achieved. Due to its nanoscrystalline grains, it is potential to obtain high coercivity for HD Nd-Fe-B magnets. However, the coercivity of HD Nd-Fe-B magnets is far less than the theoretical value even though the grain sizes are so-called nanoscale. This may be caused by the disadvantages which limits the enhancement of coercivity for hot-deformed Nd-Fe-B magnets, such as ferromagnetic intergranular phases, platelet shaped grains and surface defects et al. Reversal domains would be nucleated at ferromagnetic intergranular phases and grains' surface defects followed by propagating into $Nd_2Fe_{14}B$ grains. Besides, the platelet-shaped grains may also be the primary explanation for the relatively low coercivity which would largely increase the effective demagnetization factor, $N_{eff}$.

3.2 Coercivity enhancement of hot-deformed Nd-Fe-B magnets

Although these disadvantages could not be avoided in HD Nd-Fe-B magnets, many efforts have been made to increase the coercivity and the coercivity enhancement mechanism is also discussed.

Additions of some elements are beneficial for enhancement of coercivity in HD Nd-Fe-B magnets [31, 38-39]. It is commonly assumed that the enhancement of coercivity is contributed to modification of the grain-boundary phases by element additions. Besides, microstructure optimization and enhancement of magnetocrystalline anisotropy are both beneficial for increase of coercivity. Due to the resistance to oxide and convenient operation, rare earth fluorides have been utilized to produce 45EH and 50UH grade sintered-Nd-Fe-B magnets in practical production. In HD Nd-Fe-B magnets, the rare earth fluorides are also introduced to prepare high coercivity magnets. Xu Tang et al. [40] and Sawatzki *et al.* [41] have independently reported $DyF_3$ compounds to produce the hot compact precursors followed by hot-deformation. The substitution of Dy for part Nd which results in the formation of $(Nd, Dy)_2Fe_{14}B$ is considered to be the main reason for the enhancement of coercivity. However, the existence of a mass of $DyFe_3$ compounds in the diffusion regions between adjacent ribbons leads to large wastages of rare earth resources and decrease in remanence. Moreover, the crisis of heavy rare earth (HRE) pushes researchers to search more effective way to increase coercivity without or with a little usage of HRE. Microstructure optimization could be more worthy way to increase coercivity. In recent work, Zheng *et al.* [42] found that the coercivity can be increased from15.6 to 17.75 kOe by introducing high melting WC alloy to inhibit the growth of coarse grains. Fig. shows the SEM-BSE images of the HD samples without and with WC addition. It can be seen from Fig. 14(d) that the coarse grains are refined by the addition of WC compounds compared to the initial hot-deformed Nd-Fe-B magnets as shown in Fig.14(c). The inhibition of coarse grains growth may open a brand new door to develop high coercivity Nd-Fe-B magnets.

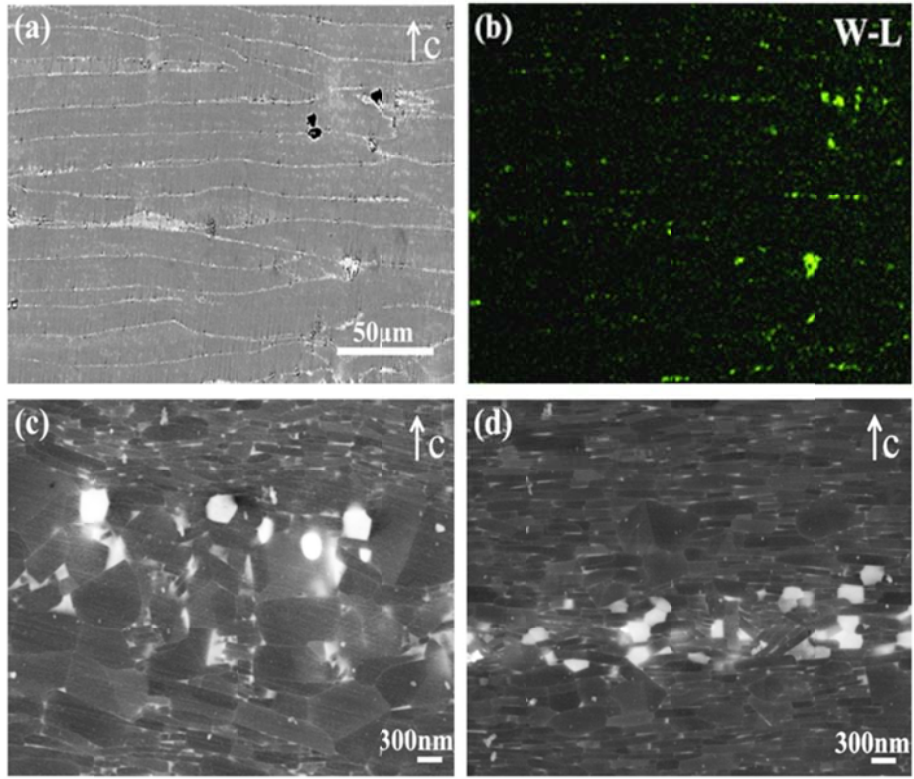

Fig. 14. (a) BSE images and (b) the EDS elemental map of W obtained from the 1.0WC sample; (c) and (d) BSE images of coarse grains region for the initial and 1.0WC sample, respectively [42].

The grain refinement and optimization of grain boundaries which would eliminate the ferromagnetic phases and surface defects, for the moment, are the most effective ways to increase coercivity. It is well accepted that the coercivity of Nd-Fe-B magnets is influenced by the content of Nd-rich phases, and the coercivity increases along with the increase of Nd-rich phases contents [13]. Recently, grain boundary diffusion of rare earth eutectic alloys in nanostructure magnets has been studied widely to increase coercivity. HDDR powders with a coercivity of 19.5 kOe are obtained by diffusion process using Nd-Cu eutectic alloy [43]. The thickened grain boundary of the HDDR powders is considered to be the primary reason for the enhancement of coercivity. The coercivity of melt-spun Nd-Fe-B powders is enhanced to 2.6 T by the Nd-Cu infiltration as well [44]. Sepehri-Amin *et al.* reported that high coercivity of ultrafine-grained hot-deformed Nd-Fe-B magnets with 2.3 T is obtained by Nd-Cu GBDP [45]. The analysis of element distribution proves that the ferromagnetic elements in the intergranular phases reduce from ~55 at.% to an undetectable level, which results in the decoupling of Nd-Fe-B grains and the increase of coercivity.

In addition, a much higher coercivity with 2.6 T is simultaneously achieved by low-temperature grain boundary diffusion of $Nd_{60}Dy_{20}Cu_{20}$ eutectic alloy and $Pr_{70}Cu_{30}$ alloy, respectively [46-48]. Researchers found that the composition of ferromagnetic elements in intergranular phases parallel to c-axis reduces substantially and thin Dy-rich shell surrounds $Nd_2Fe_{14}B$ grains. Based on the previous investigations, an optimized composition of PrNd-Cu alloy is applied and the HD

Nd-Fe-B magnet with higher coercivity of 2.7 T is obtained [49]. The BSE SEM images of the HD and diffusion-processed magnets are illustrated in Fig. 15. The images of the diffusion-processed magnets with the c-axes in-plane and out-of-plane are shown in Fig. 15(b) and (c), respectively. The Nd-rich intergranular phases become thicker after eutectic alloy diffusion treatment.

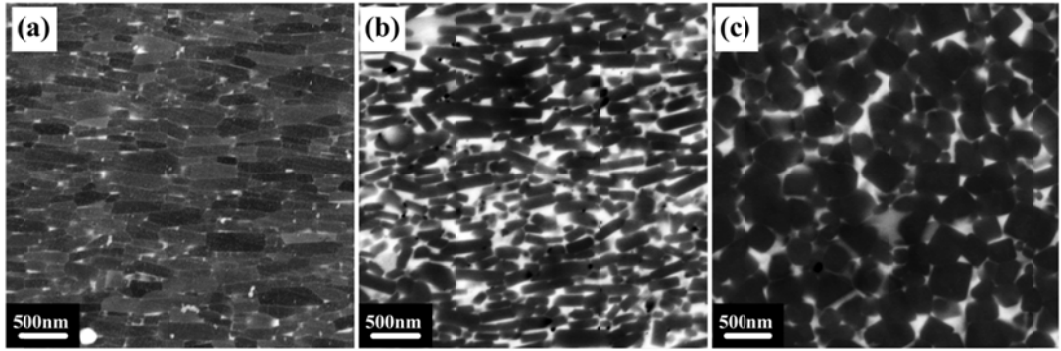

Fig. 15. TEM-EDS images of the hot-deformed and diffusion-processed magnets.

Based on the demagnetization profiles in Ref. 52, the coercivity of the surface regions is stronger than that of center regions of the diffusion-processed magnets. Besides, the TEM images in Ref. 52 indicate the diffusion process is actually inhomogeneous which results in the magnetization behaviors similar to nanocomposite magnets. Recently, the coercivity has achieved the highest value of 28 kOe without HRE addition by diffusing the low melting temperature glass forming $Pr_{60}Al_{10}Ni_{10}Cu_{20}$ alloys [50]. Magnetic isolation of the $Nd_2Fe_{14}B$ grains by the Nd-rich amorphous/crystalline intergranular phases, which are considered to be nonmagnetic phases, is attributed to the large enhancement of coercivity. Bance *et al.* [51] reported that the thickness of surface defects, where are commonly considered to possess lower magnetocrystalline anisotropy, plays a critical role in determining the magnetization reversal process. Besides, Ref. 51 reveals that the single-grain computed results of angular dependence of the coercive field appear to match well with experimental behavior when imperfect grain alignment of real magnets is taken into account. Anyway, the enhancement of coercivity is mostly resulted from the modification of grain boundaries by rare earth eutectic alloys infiltration process.

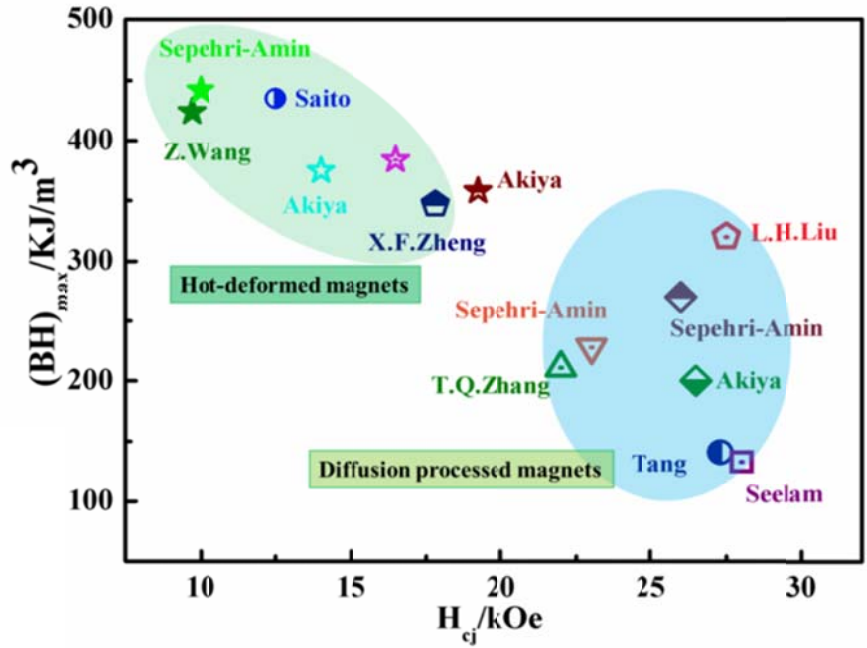

Fig. 16. Current development of coercivity in HD Nd-Fe-B magnets.

Fig. 16Fig. shows the summarization of the current developing situation of coercivity in HD Nd-Fe-B magnets according to some references. It can be seen from the summarization that the coercivity can be largely enhanced by appropriate optimization diffusion process. However, simultaneously, the remanence is dramatically decreased. Generally, the remanence ($M_r$) could be described as follows:

$$M_r = A(1-\beta)\frac{d}{d_0}\overline{\cos\theta}M_s$$

Where $A$ is the volume fraction of the positive domain parallel to the pre-magnetization, $(1-\beta)$ the volume fraction of hard magnetic phase, $d_0$ the theoretical density, $d$ the actual density, $\overline{\cos\theta}$ the orientation degree and $M_s$ the saturation magnetization of magnets. After diffusion treatment, the excessive nonmagnetic phases will increase the value of β resulting in the remanence decrease. In fact, the deterioration of grain orientation, which may be an ignored key factor, decreases the remanence and energy production as well. In order to achieve high coercivity without sacrifice of remanence, Akiya *et al.* [52] introduces an effective method of low eutectic Nd-Cu diffusion process under expansion constraint. However, the enhancement of coercivity is limited.

In sintered-Nd-Fe-B magnets, an optimized structure that $(Nd,Dy)_2Fe_{14}B$ or $(Nd,Tb)_2Fe_{14}B$ surrounds $Nd_2Fe_{14}B$ phases can be formed by GBDP to results in dramatic increase of coercivity without evident sacrifice of remanence. In hot-deformed Nd-Fe-B magnets, high coercivity may also be obtained if such core-shell structure is obtained. Liu *et al.* [53] reported a diffusion-processed HD magnet with a coercivity value of 2.75 T and a remanence value of 1.3 T using

Nd$_{62}$Dy$_{20}$Al$_{18}$ alloys as a diffusate. In this study, both the core-shell structure with high magnetocrystalline anisotropy and magnetic isolation decoupled by nonmagnetic intergranular phases result in the large enhancement of coercivity with just a little decrease of remanence. Recently, Wang *et al.* [54] developed a route to obtain excellent comprehensive magnetic properties of HD Nd-Fe-B magnets by Dy-Cu press injection and subsequent annealing. The demagnetization curves and magnetic properties of the hot-deformed and Dy-Cu press injected magnets in different stages of the procedures are shown in Fig. . High performance can be achieved by optimized annealing while the near-surface microstructure is remarkably improved and core-shell structure is formed by Dy-Cu press injection.

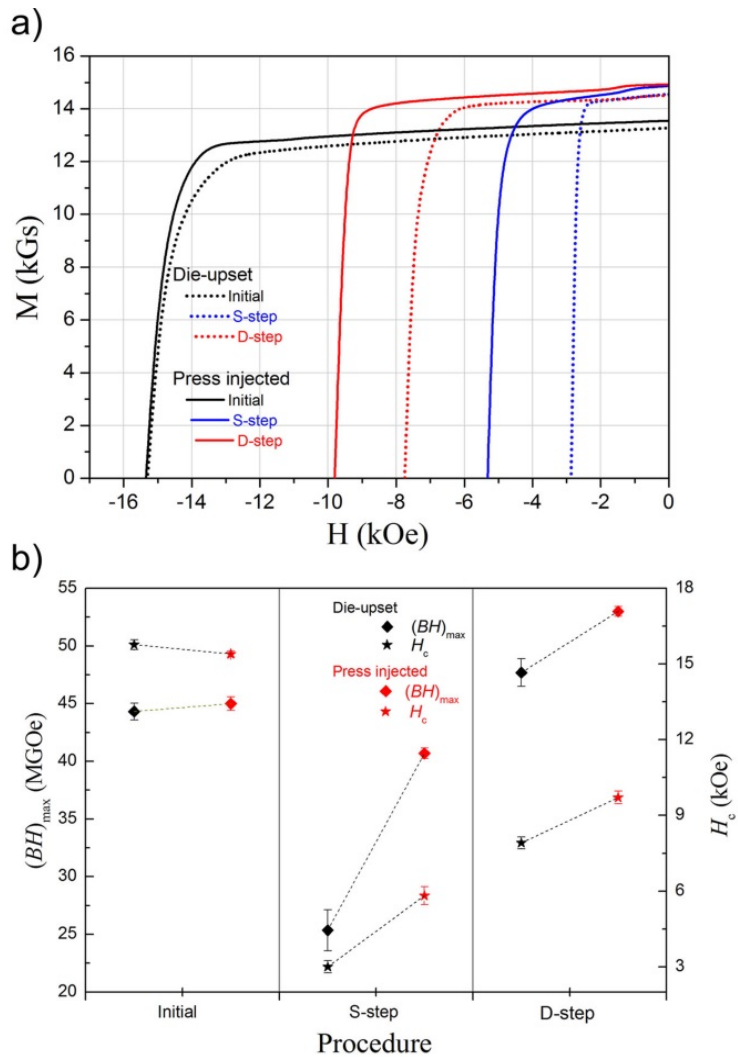

Fig. 17. (a) Demagnetization curves of the hot-deformed and Dy-Cu press injected magnets, (b) Magnetic properties of the hot-deformed and Dy-Cu press injected magnets in different stages of the procedures: initial state, S-step and D-step annealed states [54].

3.3 Discussion about coercivity enhancement of HD Nd-Fe-B magnets

Based on the discussion above, the coercivity is closely related with facet defects of grains and magnetism of intergranular phases. High coercivity could be obtained by modification of grain boundary and intergranular phases. Sepehri-Amin *et al.* [45] investigated the influence of magnetism of intergranular phases on coercivity by finite element model as shown in Fig. . The micromagnetic simulation result indicates that the reduction of the magnetization in the intergranular phases would effectively improve coercivity, which is in agreement with the experimental results.

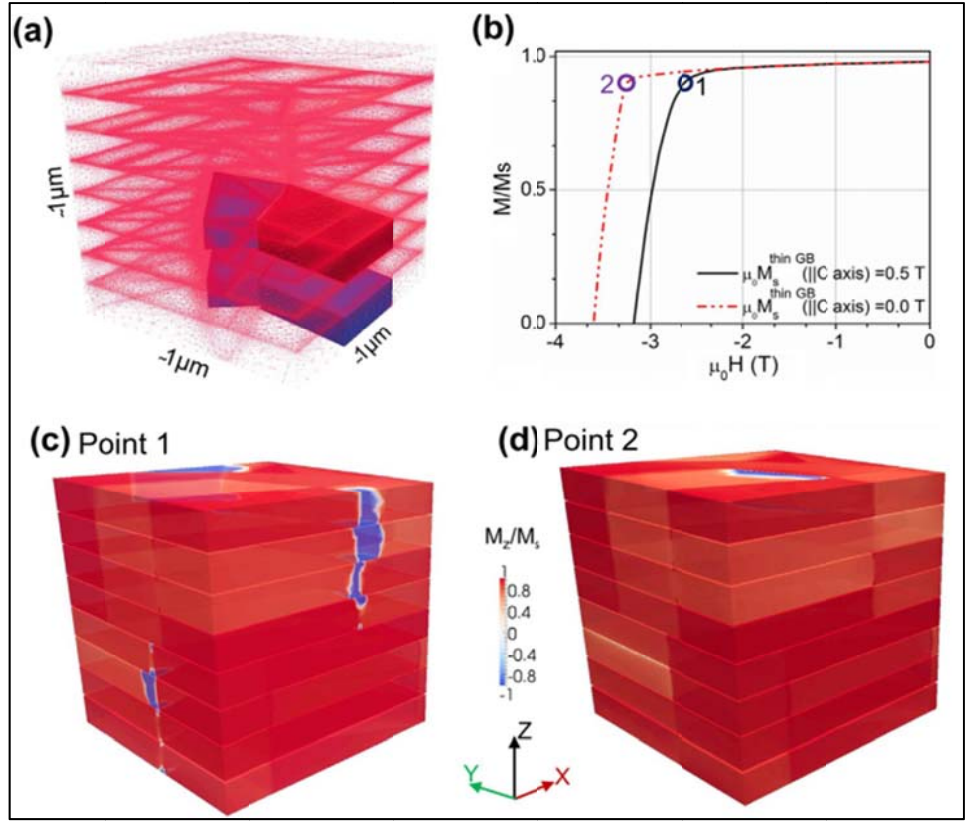

Fig. 18. Micromagnetic simulations of the intergranular phases by finite element model [45].

Besides, according to the S-W model, the magnetic isolation by the nonferromagnetic intergranular phases is also beneficial for the enhancement of coercivity in HD Nd-Fe-B magnets. Bance *et al.* [51] studied the influence of defect thickness on the angular of coercivity, the results indicate that with increasing defect thickness the magnetization reversal mechanism shifts from nucleation to de-pinning. Tang *et al.* [49] analyzed the magnetization behaviors of the diffusion processed HD magnets and found an S-shaped magnetization profile as shown in Fig. 18 suggested a strong pinning behavior. From Fig.19, the magnetization shows a 3-steps process. Domain wall reversible motion within $Nd_2Fe_{14}B$ grains is dominant in the first step while domain wall pinning and magnetization reversal of grains are primary magnetization process in second step. As a result, the coercivity may be dramatically enhanced if the grains are perfectly magnetic isolated by non-ferromagnetic phases and the coherent rotation could be achieved. However, Tang et al. also reported that the distribution of low eutectic alloys is actually inhomogeneous, some grains are surrounded by infiltrated phases while some grains are still stacked together. This may be because the distribution of grain boundaries is inhomogeneous in initial HD

Nd-Fe-B magnets. The inhomogeneous distribution of Nd-rich phases would result in open recoil loops in Fig. 15 (a) which are not beneficial for practical utilization for diffusion processed magnets.

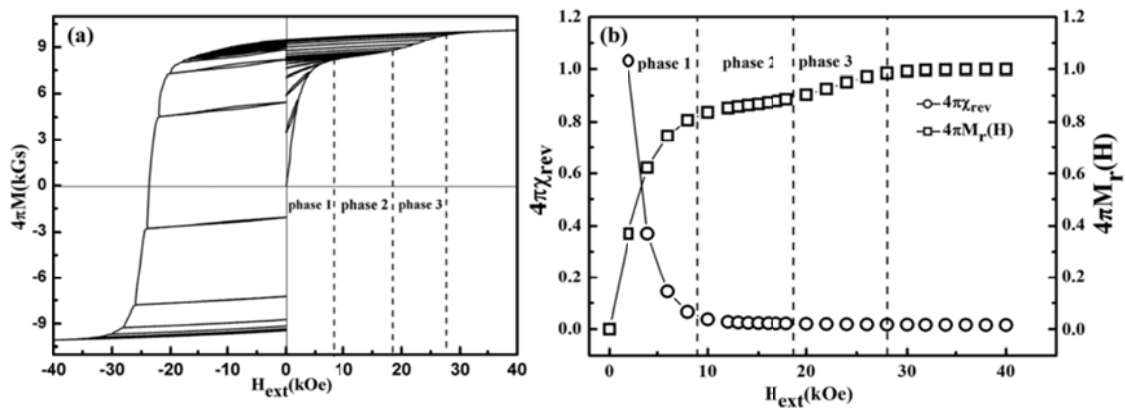

Fig. 19. (a) Recoil loops of the diffusion processed sample; (b) field dependence of remanence and reversible susceptibility [49].

In summary, the coercivity has become the dominant obstacle for extensive application of Nd-Fe-B magnets. Researchers have to face the contradiction that how to achieve high coercivity (HRE-free) without decrease of remanence. Based on the previous investigations, high coercivity may be obtained by appropriate optimizations of grain boundaries and intergranular phases. Firstly, the Nd-rich intergranular phases which are commonly considered as nonmagnetic phases should be noted as the key microstructure for the coercivity enhancement of Nd-Fe-B magnets. If the ferromagnetism of the intergranular phases could be weaken and the neighbor grains could be decoupled, high coercivity would be achievable; secondly, core-shell structure, which could inhibit the reversal domain nucleation due to the magnetic hardness of grain surfaces, may also be an effective way to obtain high coercivity. High coercivity without reduction of remanence has been achieved in thin films which has almost ideal microstructures [55, 56]. In any case, it will be great challenges for which researchers have to struggle to achieve higher coercivity than that of thin films.

**Section 4 HD nanocomposite Nd-Fe-B magnets**

Recently, new rare-earth permanent magnetic materials with high magnetic properties have attracted much interest due to the increasing demand for environment-friendly and energy-saving industry. However, the scarcity and high cost of rare earth (RE) limit the applications of high-performance rare-earth compounds, like Nd-Fe-B permanent magnets. The relatively high concentration of iron (cobalt) atom of nanocomposite magnets reduces the usage of RE elements, making great contributions to the sustainable development of RE industry. Therefore developing the nanocomposite magnet has become one of the main ways for easing RE shortage. In addition to the reduced RE content, a high soft-magnetic phase fraction in nanocomposite magnets also contributes to the large saturation magnetization for potentially high energy products. Predication implies that the exchange interaction between "soft-hard" magnetic phases can perform well if the "soft-magnetic" phase is thinner than the domain wall thickness, only 3.9 nm for $Nd_2Fe_{14}B$ system. The magnetostatic interaction is demonstrated to relax the strict requirements for the size of the "soft-magnetic" inclusions. To date, numerous "soft-hard" composite systems have been investigated. Current the bonded Nd-Fe-B nanocomposite magnets fabricated by Nd-Fe-B/α-Fe magnetic powders have been widely used. However, the isotropic magnetic characterization and high concentration of nonmagnetic resin adhesive largely deteriorate the magnetic performance. The obtained maximum energy product $(BH)_{max}$ of the isotropic powders lower than 16 MGOe. Hence, how to fabricate high-anisotropy nanocomposite magnets has become a significant issue for the industry and academia [57].

The hot-pressed/hot-deformed technique has become an important branching for preparing fully dense anisotropic nanocrystalline magnets. Generally, we can obtain anisotropy hot-deformed Nd-Fe-B composite magnets by different pathways. The nanocomposite Nd-Fe-B magnets were prepared with isotropic Nd-lean magnetic powders. Compared with the Nd-rich magnetic powders, the Nd-lean magnetic powders give rise to the degradation of deformability for the bulks. It is difficult to realize stress-induced crystallographic orientation in single nanocomposite Nd-Fe-B magnets. Subsequently, the co-doping technique has been widely used to improve the deformability of nanocompsite magnets, such as Nd-rich/Nd-lean [58, 59], Nd-rich/Fe(Co) systems [60]. As low-melting RE-rich intergranular phase plays an essential role in the stress-induced solution-precipitation creep process [61-63], RE-rich eutectic can be diffused into the grain boundaries to facilitate the formation of highly textured structure.

4.1. Nanocompsite magnets prepared by Nd-lean powders

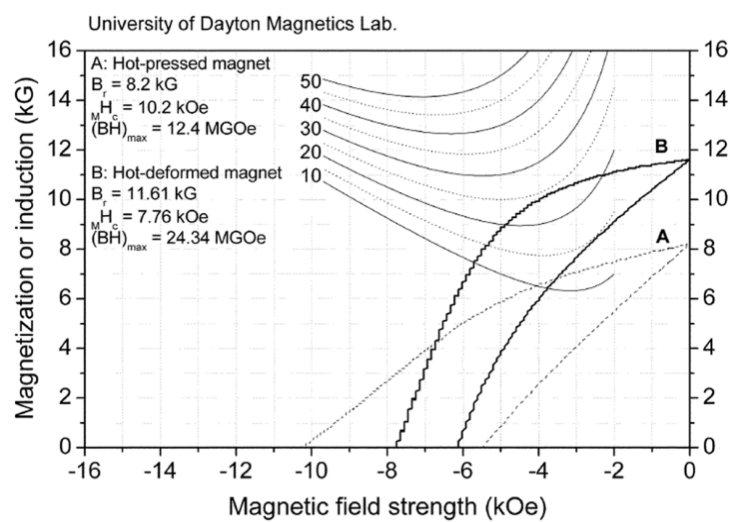

Fig.21. Demagnetization curves of rare-earth-lean Nd-Fe-B nanocomposite magnets parallel and perpendicular to easy-magnetization direction [64].

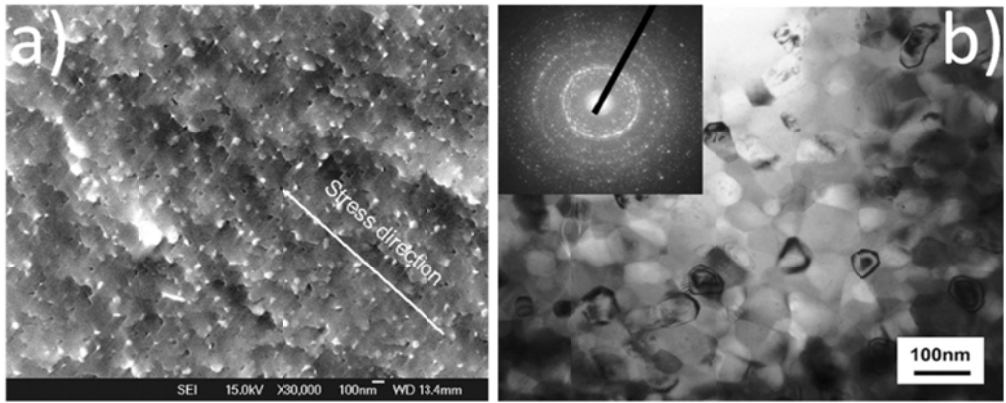

Fig.22. (a) SEM fracture surface of hot-deformed magnets with a composition of $Nd_{9.3}Pr_1Dy_{0.3}Fe_{77.4}Co_{6.1}Ga_{0.2}B_{5.7}$ (at.%); (b) TEM micrograph and SAED pattern (inset) of magnets with a composition of $Nd_{9.3}Pr_1Dy_{0.3}Fe_{77.4}Co_{6.1}Ga_{0.2}B_{5.7}$ (at.%) [64].

To achieve the Nd-lean magnetic powders with the rare-earth contents lower than the chemical stoichiometry, the ingots with lower RE contents are usually quenched into melt-spun ribbons [64-67]. And then, the nanocomposite magnets fabricated by these crystallized Nd-lean magnetic powders are obtained using the hot-pressed and hot-deformed processes. The demagnetization curves of nanocomposite Nd-Fe-B hot-deformed magnets (nominal composition: $Nd_{9.2}Pr_1Dy_{0.3}Fe_{77.3}Al_{0.2}Co_{6.1}Ga_{0.2}B_{5.7}$ at. %) are measured parallel and perpendicular to easy-magnetization direction (Fig.21). It can be found that the nanocomposite magnets exhibit a weaker anisotropy because of the low [001]-oriented structure. Microstructure characterization indicates that the short-axes of nanocrystals are preliminarily aligned along the loading direction (Fig. 22). However, a large number of nanocrystals show equiaxed morphology. Through increasing the deformation temperature (i.e. 900 °C), more $Nd_2Fe_{14}B$ nanocrystals could be shortened along the easy-axis and elongate along its perpendicular direction [68]. In view of lacking Nd-rich phases, the textured structure is probably derived from shear deformation under the combination action of pressure and temperature. Specifically, the anisotropic elasticity modulus of $Nd_2Fe_{14}B$ crystal accelerates the rotating and sliding of lattice plane. Particularly note that the magnetic properties of nanocomposite magnets decreases with increasing the content of "soft magnetic" phases. Up to now, the maximum energy products obtained in laboratory are only in the range of 20–25 MGOe. Apparently, the obtained unsatisfactory remanence and maximum energy product is due to the disordered crystallographic orientation. Improving the grain alignment in nanocomposite magnets has become a challenge for developing the high-performance RE permanent magnets. To release the awkward situation, the uniaxial ultrahigh stress is employed to induce the orientation of Nd-lean amorphous magnetic powders. This method has successfully yielded the highly textured structures [65-66, 69-70].

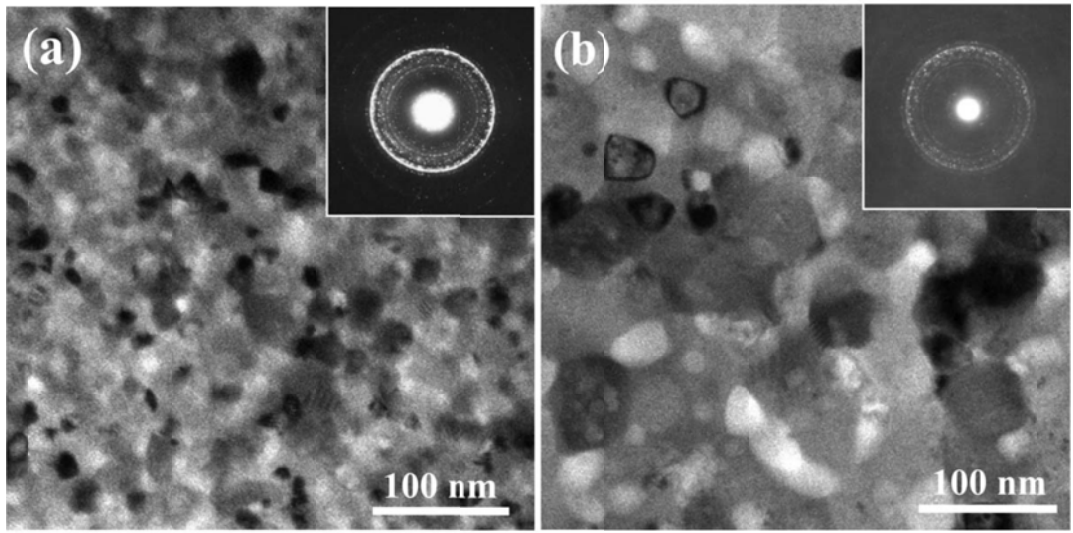

Fig. 23. TEM images and electron diffraction SAED patterns of amorphous $Nd_9Fe_{85}B_6$ alloy (a) subjected to HPTD and subsequently annealed at 600 °C for 10 min and (b) directly annealed at 725 °C for 10 min [69].

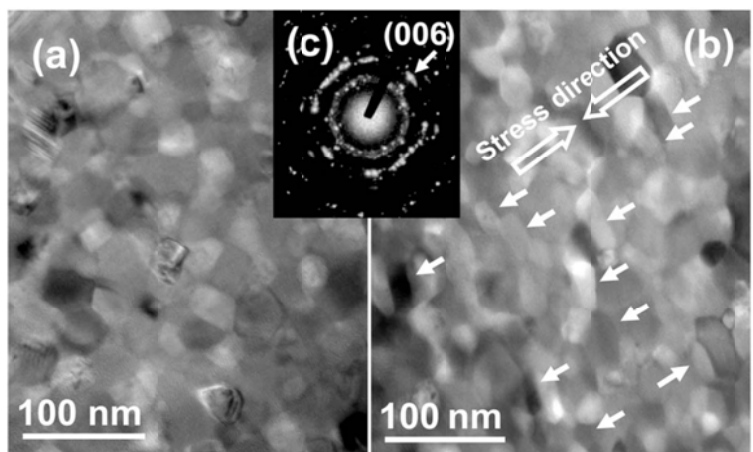

Fig. 24. TEM bright-field images (a) perpendicular and (b) parallel to the stress direction (c) SAED patterns of the bulk amorphous $Nd_9Fe_{85}B$ after hot deformation [66].

In amorphous system, the crystallization control in amorphous alloys is of particular importance for the development of high-performance nanocrystalline materials. Li et al.[79] succeeded in controlling α-Fe and $Nd_2Fe_{14}B$ crystallization processes in amorphous $Nd_9Fe_{85}B_6$ by a combination of severe plastic deformation at room temperature and subsequent heat treatment, as shown in Fig. 23. The $Nd_2Fe_{14}B$/α-Fe nanocomposite magnets prepared by this approach possess homogeneous microstructure with a small size, 15 nm for α-Fe phase and 26 nm for $Nd_2Fe_{14}B$ phase, and therefore show enhanced magnetic properties as compared to those prepared by directly annealing amorphous $Nd_9Fe_{85}B_6$. The small starting phase size gets ready for the strong "hard-soft" coupling in ultimate bulk magnets. To obtain a strong [001]-oriented texture, Nd-lean amorphous powders are hot-pressed and hot-deformed at a large uniaxial stress of 310 MPa. The unusual texture formation is attributed to a preferential nucleation of $Nd_2Fe_{14}B$ crystals in amorphous matrix induced by ultrahigh stress. Fig. 24 shows the platelet-shaped microstructure, in which $Nd_2Fe_{14}B$ nanocrystals with sizes of 35 - 50 nm in diameter and 18 - 25 nm in thickness can be observed (Fig.24). Selected area electron diffraction (SAED) reveals that the polycrystalline diffraction signals are due to weak [001] texture caused by α-Fe phases. Due to the hash ultrahigh stress, other simple techniques are urgently needed to develop the high-anisotropy nanocomposite magnets.

4.2 Nanocompsite magnets prepared by the co-doping technique.

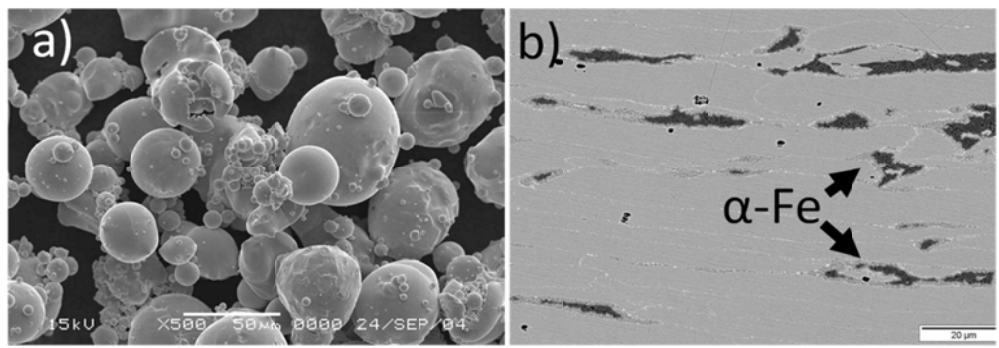

Fig. 25. (a) SEM micrograph of Fe-Co powder particles; (b) SEM back scattered electron image of a hot-deformed composite $Nd_{13.5}Fe_{80}Ga_{0.5}B_6$/α-Fe (91.7 wt.%/8.3 wt.%) magnet [71].

As Nd-lean magnetic powders own a poor deformability, co-doping techniques have been developed to fabricate high-performance nanocomposite magnets. For instance, Nd-rich magnetic powders are co-doped with Nd-lean magnetic powders to improve the deformability of bulk samples. Experimental results have demonstrated that the maximum energy product of co-doped magnets can increase to 45 MGOe [71]. It indicates that the Nd elements diffuse from the Nd-rich magnetic powders to Nd-lean magnetic powders to improve the crystallographic orientation of Nd-lean structure. Therefore, this method realizes the [001]-oriented texture in nanocomposite magnets. Besides the Nd-rich magnetic powders, the "soft magnetic" particles, like Fe-Co phase (Fig. 25(a)), are also blended with the Nd-rich powders, and then hot deformed into bulk nanocomposite magnets. BSE SEM image in Fig. 25(b) shows that the dark gray α-Fe phases are embedded into the brighter Nd-Fe-B matrix. In contrast to the nanocompsite structure, the α-Fe phase with a size of 20 μm is far beyond the limitation for the short-range exchange coupling interaction. Apparently, the long-range magnetic interaction should play a critical role in the improved magnetic properties [60]. To further disperse the soft magnetic phases, the soft phase can be deposited on the Nd-rich powder surfaces. This method effectively increases the maximum energy product of hot-deformed magnets, up to the maximal value of 55 MGOe.

4.3 Rare-earth-rich liquid enhancing the deformability of nanocomposite magnets

As the formation of texture in nanocomposite magnets is difficult, it is of importance to reveal the mechanism of texture formation in hot deformation. Prior researches indicate that RE-rich phase is an essential factor for magnet's deformability. To investigate the deformation mechanism of rare-earth-lean nanocomposite magnets, the low-melting RE-rich eutectic alloy is added into the hot deformation process. Thus the formation of textured structure can be controlled by adjusting the chemistry composition and concentration.

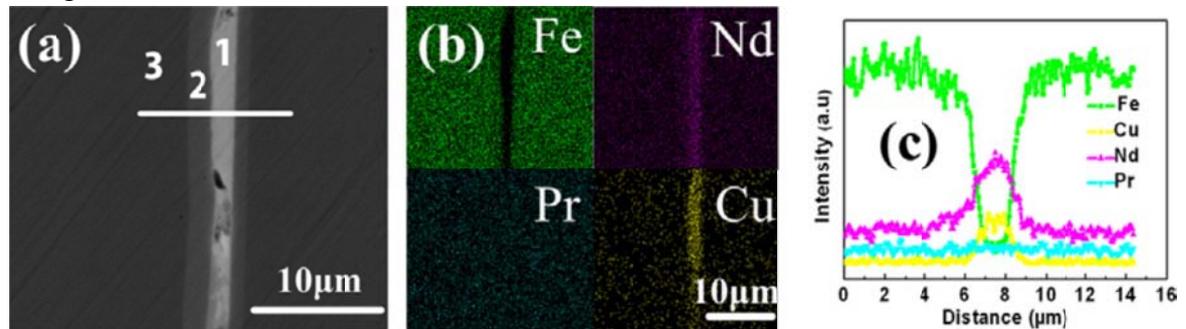

Fig. 27. (a) BSE SEM image showing the interface of particles in the sample hot pressed at 600 °C, (b) EDS maps of elements in this area and (c) EDS line scan across the interface region marked in (a) [72].

Fig. 27(a) shows the microstructure of isotropic precursor hot-pressed at 600 °C, where two adjacent powders and a narrow boundary area between the two powder particles is observed. In terms of imaging contrast, there are at least three regions can be captured: (1) bright region; (2) grey region; and (3) dark region. The energy-dispersive X-ray spectrometry (EDS) analysis indicates that the bright region is the Nd-Cu-rich phase and the dark region corresponds to the raw nanocomposite particle without Nd-Cu diffusion; the presence of the transition grey region suggests that Nd-Cu diffusion takes place in this area. This result indicates that the Nd-Cu eutectic diffusion process starts during the hot-pressing process.

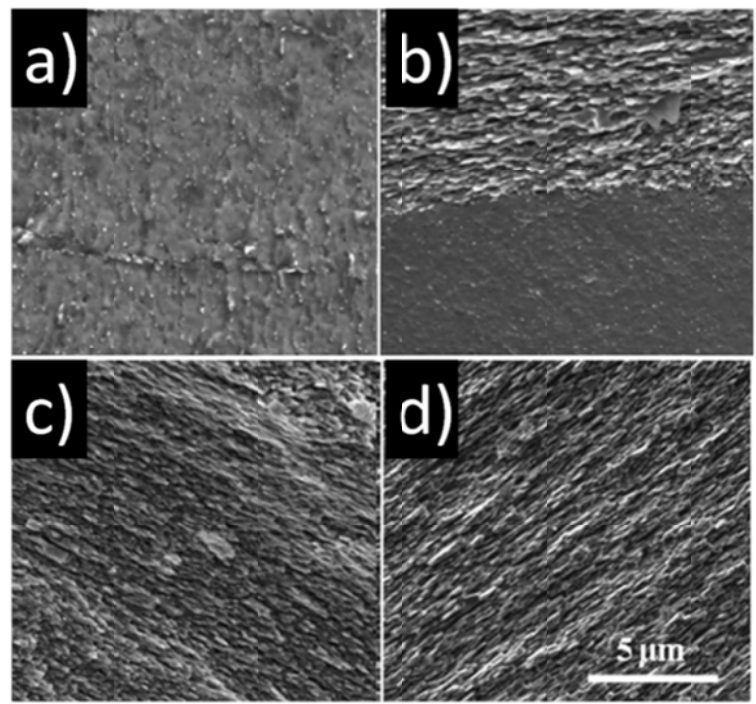

Fig.28. Field emission SEM micrographs of hot-deformed nanocomposite magnets

with various Nd-Cu alloy addition. x = 0 (a), 4*wt*.% (b), 6*wt*.% (c), and10 *wt*.% (d) [73].

Besdies the hot-pressed process, the mechanism of microstructure evolution in eutectic-doped hot-deformed magnets is also studied. The effect of low-melting Nd-rich intergranular phases on the formation of platelet-shaped grain and textured structure is critical. Tang *et al.* [73] studied the the effect of Nd-rich phases on the texture formation in nanocomposite magnets by diffusing Nd-Cu eutectic. The results demonstrate that the Nd-rich intergranular phases are in favor to the anisotropy of Nd-Fe-B magnets, as shown in Fig.28. With adding appropriate Nd-Cu content, the well-aligned platelet-shaped structure can be obtained.

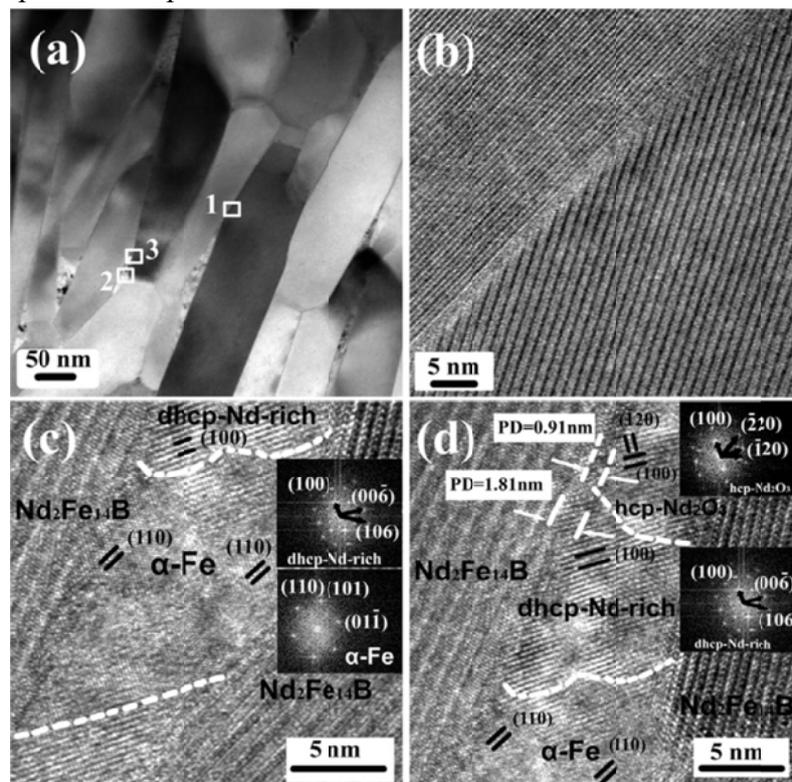

Fig.29. TEM/HRTEM and fast Fourier transform images of hot-deformed nanocomposite magnets with the deformation rate of 30% [72].

In addition, the crystal structures in the interganular phases are correlated with the chemical composition (Fig. 29). Tang *et al.* [72] studies the microstructure of intergranular phases in nanocomposite magnets, as shown in Fig.29. HRTEM image of the grain boundary located at position 1 marked in Fig. 29(a) indicates an evident intergranular phase. The fast Fourier transform (FFT) patterns taken in corresponding areas show that the intergranular phases consisted of dhcp-Nd-rich phase and hexagonal close-packed (hcp) $Nd_2O_3$ phase with lattice parameters of a = 3.83 Å and c = 5.99 Å. The decomposition from the face-centered cubic $NdO_x$ (fcc-$NdO_x$) phase to hcp-$Nd_2O_3$ and dhcp-Nd-rich phases occurs at the temperatures below 863 °C. In this work, the relatively low hot deformation temperature 850 °C creates favorable conditions for this decomposition. Furthermore, the dhcp-Nd-rich phase transforms easily into hcp-$Nd_2O_3$ through oxidation, because the latter is the most stable phase and formation of hcp-$Nd_2O_3$ is energetically favored. After adding Nd-Cu alloy, the coercivity of hot-pressed magnets is improved largely, implying that the Nd-Cu liquid starts diffusion into powders during hot-pressing process. Therefore the subsequent high-temperature deformation is beneficial to the diffusion and texture formation during hot deformation. The above findings reveal that the RE-rich intergranular phase play a vital role in facilitating the formation of highly textured structure. The lack of RE-rich phases is the main reason for the poor deformability of nanocomposite magnets.

**Section 5 Radially-oriented Nd-Fe-B ring magnet prepared by backward extrusion method**

Radially-oriented Nd-Fe-B ring magnet (RM) is a special Nd-Fe-B magnet with ring shape and the easy *c*-axis along the radial direction (Fig.30). Because of the RM is convenient for magnetization and assembly, It has great potential applications especially in many kinds of motors for its special shape. So far, there have been three kinds of Nd-Fe-B ring magnets (RMs): bonded RM, sintered RM and hot-deformed RM. Bonded PM is isotropic and possesses lower energy density than the other two RMs. Sintered RM experiences a complex preparation that costs very long time. In contrast, hot-deformed RM can be prepared via a facile near-net-shape method, which shows a significantly higher efficiency. Therefore, more and more attention has been focused on hot-deformed RM. As a unique hot deformation technique, backward extrusion is commonly used for preparation of radially oriented Nd-Fe-B ring magnet (RM). Besides, backward-extruded RM has high power-to-volume ratio, good corrosion resistance and lower rare earth content. Therefore, it has great potential applications in servo motors, EPS motors and driving motors.

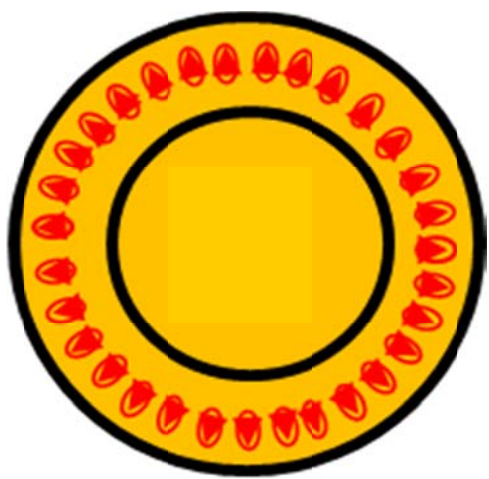

Fig. 30. Schematic illustration of radially-oriented Nd-Fe-B ring magnet

The procedure of backward extrusion method is illustrated in Fig. 31. It consists of two steps: hot-pressing and hot deformation (back extrusion). The HP step is employed to densify the melt-spun powders at a certain temperature (600-700°C) under pressure. As melted Nd-rich phase existed in the Nd-Fe-B magnet at HP temperature, it is easy to obtain magnet with nearly full density. HP magnet mainly consists of equiaxed $Nd_2Fe_{14}B$ grains and thus exhibits isotropic magnetic performance. HD step aims for ring shaping and texture formation at HD temperature (700-850°C) in the extrusion process.

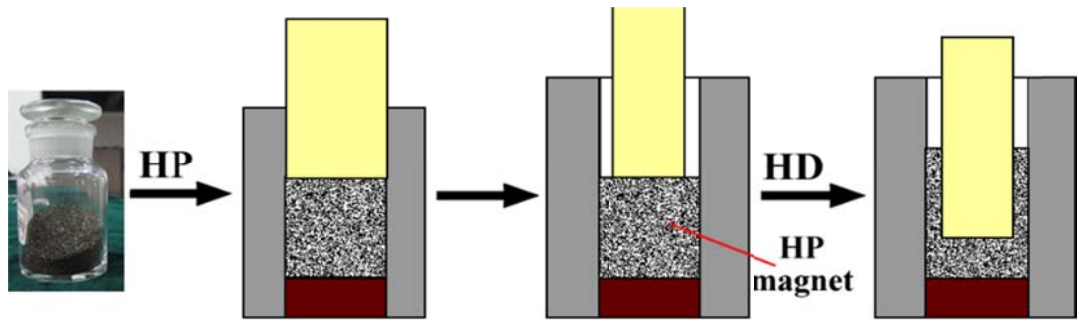

Fig. 31. Schematic illustration on the procedure of backward extrusion method

5.1 Texture evolution in backward extrusion of Nd-Fe-B ring magnet

The HD RM possesses bright prospect in brushless DC motors [74]. But some researches indicate that the inhomogeneity of magnetic performance limits the HD RM applications in industry to some extent. For example, remanence decreased quasi-linearly from the inner to outer surface of backward-extruded (BE) RM along radial direction [75-76]. In axial direction, remanence increased first and then decreased from the top to bottom of RM [77-78]. To disclose the reasons for such inhomogeneity, the mechanisms of texturing and material flow have to be clarified.

Up to date, three main mechanisms have been proposed for explaining the formation of platelet $Nd_2Fe_{14}B$ grains in hot-deformed magnet. Li *et al.* suggested the mechanism based on dissolutions of non-favourably oriented crystallites in the Nd-rich phase and their re-precipitation on the crystallites, which have their *c*-axes oriented parallel to the stress direction [79]. An extensive study by Grünberer *et al.* suggested a dominating role of the diffusion creep mechanism, which is accompanied by rheological and viscous flow [75]. Yuri and Leonowicz have complemented another mechanism, which considered the mechanical rotation of grains when they deformed plastically and the alignment results from geometrical anisotropy of the crystallites [80-81]. However, the subsequent evolution of these $Nd_2Fe_{14}B$ platelets was rarely reported because it greatly depends on the preparation strategy.

Yin et al. [5] have done a systematic investigation on the texture evolution and materials flow over partly-deformed ring magnet. Partly-deformed ring magnet was divided into two parts: ring-shaped part and cylindrical part. Their investigation reveals that the formation of the platelet-shaped grains merely takes place in cylindrical part and the depth of texture zone is limited from the bottom of punch. Further investigation suggests that equiaxed grains are distributed in the bottom region in cylindrical part because no material flow happened there (Fig. 32). Moreover, orientation of $Nd_2Fe_{14}B$ platelets is changed in the region where ring-shaped part and cylindrical part meet (Fig.33), resulting in 90° rotation of *c* axis when $Nd_2Fe_{14}B$ platelets flows from cylindrical part into ring-shaped part.

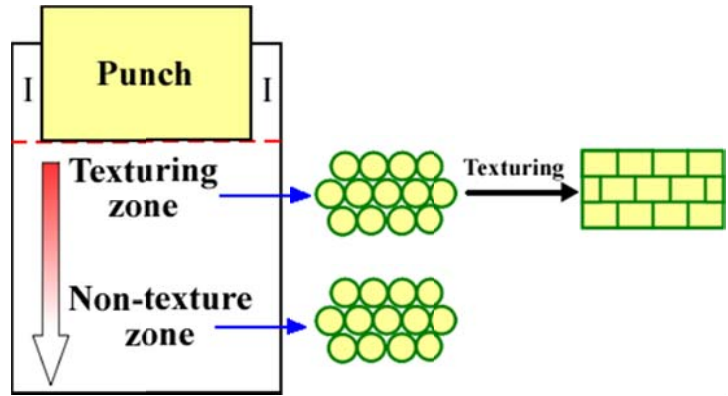

Fig. 32. Schematic illustration on the texturing zone and non-texture zone in the cylindrical part of ring magnet prepared by backward extrusion method.

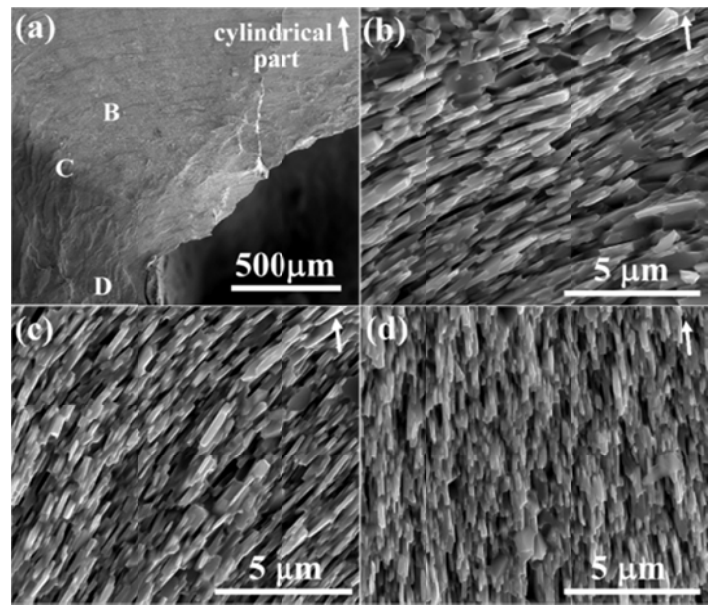

Fig. 33. SEM images of the rigion where cylindrical part meet with ring-shaped part. (a) is the image under low-magnification. (b), (c) and (d) are the high-magnification images of region B, C and D in image (a), respectively. White arrows represent the pressing direction.

5.2 Magnetic performance improvement of backward-extruded Nd-Fe-B RM

    Radially-oriented RM exhibits inhomogeneity in remanence along axial direction [76, 82]. This inhomogeneity was ascribed to the texture difference in axial direction [83, 84]. To solve this problem, Tang *et al.* [85] has developed a pre-deformation technique. The HP magnet was firstly performed a 52% deformation. Then it was backward extruded into RM. RM prepared with this technique shows significantly improvement in the homogeneity of remanence and coercivity along axial direction (Fig. 34). Moreover, the remanence at the same region of RM was also raised with the pre-deformation technique. The microstructure over regions 1, 2, 3 and 4 of the ring magnet confirms better texture in the above regions, especially region 1 (Fig.34). But the alignment direction of platelet $Nd_2Fe_{14}B$ grains shows nearly 90 degrees

difference between region 1 and region 4, which could be ascribed to the rotation of the platelet Nd$_2$Fe$_{14}$B grains during BE. Nd$_2$Fe$_{14}$B platelets in region 1 are inherited from the counterpart formed during pre-deformation process while these Nd$_2$Fe$_{14}$B platelets rotate when they flow from the cylindrical part to the ring part (Fig. 33).

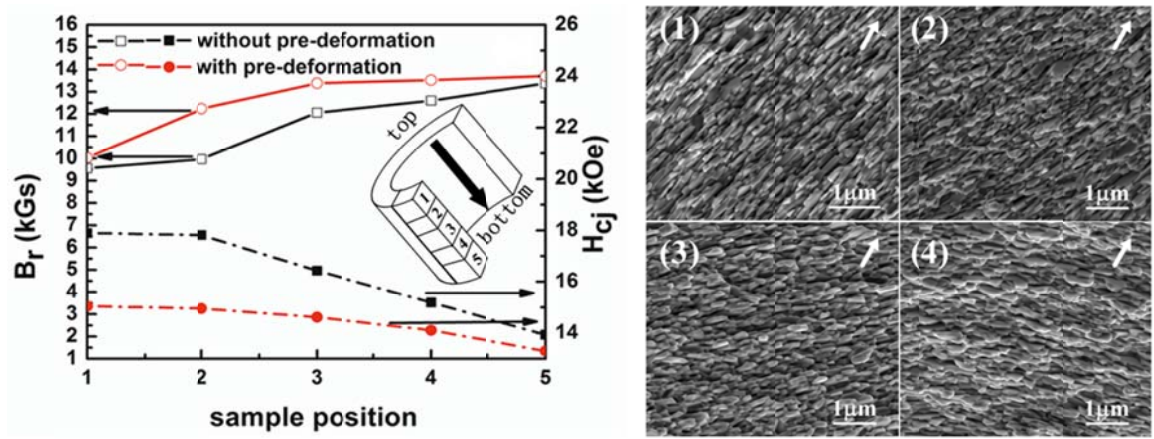

Fig. 34. Magnetic performance and the SEM images of the samples cut from different axial positions of the ring magnet prepared by pre-deformation and BE method. White arrows represent the radial direction.

Besides the improvement of remanence homogeneity, the enhancement of magnetic performance was another focus of RM research. Yi *et al.* [4] has investigated the effect of Nd content on the remanence of RM and the Nd content of 13.34 at.% was optimized. Gutfleisch *et al.* [84] found that deformation temperature and deformation speed have great influence on the coercivity while little on the remanence of the RM prepared from HDDR powder. Larger deformation rate is good for coercivity but deformation temperature has an optimum value of 725 $^o$C. The optimized deformation temperature for melt-spun powder was determined by A.H. Li et al to be 800 $^o$C for texture formation but 660$^o$C for coercivity [86]. Besides, deformation rate of 0.01 and deformation ratio of 70% was optimized for best texture and the consequent remanence.

5.3 Solution for tangential crack in Nd-Fe-B RM

Radially-oriented RM prepared by backward extrusion method usually shows cracks along tangential direction on the inside surface at its top region (Fig. 35). The region with cracks would be cut off and thus leads to the waste of Nd-Fe-B. To solve this problem, Satto *et al.* [87] had employed an extra punch to press the cracks along the axial direction until they were healed. Dirba *et al.* [88] took a similar measure: they exposed an axial pressure (not lower than 0.5MPa) on the top region of RM during backward extrusion. However, the above two works do not mentioned the reason for tangential cracks. Previously, it was reported that the cracks have something to do with the low content of Nd-rich phase [89] because the Nd-rich phase performs as glue during the material flow. Besides, Yin *et al.* [90] have noted the phenomenon that no tangential cracks existed at the middle or bottom region of RM (Fig.36). To reveal the reason, they carried out a backward extrusion over three packed samples (Fig. 37). The interface between the middle and bottom samples exhibits little change (Fig. 38(a)), which indicates that no material flow took place in the bottom sample. In other words, material flow only takes place at a limited depth below the punch during backward extrusion. The top and middle samples show strange shape and the interface between them are special (Figure 38(b)). Most of the top sample was extruded into ring magnet. But the bottom region of top sample exhibits a convex edge. Although it remains part of the top sample at this stage, it would contribute mainly to the ring part of the middle sample. The top surface of middle sample is concave-shaped, indicating the outward flow of Nd-Fe-B from the center to form ring magnet. The convex bottom of top sample should be ascribed to the slower material flow in this region due to the friction at the end surface of punch. Furthermore investigation reveals that the arc-shaped zone I (Fig. 39), which contacts with the punch, plays the key role in the elimination of tangential cracks. Guruswamy *et al.* [91] had testified the existence of Zone I and Zone II by using finite element analysis. Yin *et al.* [78] confirmed that the difference in texture was resulted from that in the flowing path. Moreover, different shapes of punches were redesigned and used for preparation of RM (Fig.40). The result shows that the half-spherical shape is the best one and could eliminate effectively the cracks at the top region of RM.

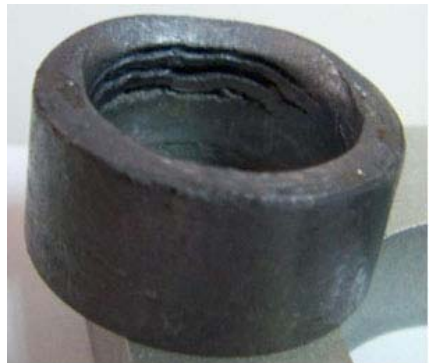

Fig. 35. Tangential cracks at the top region of ring magnet.

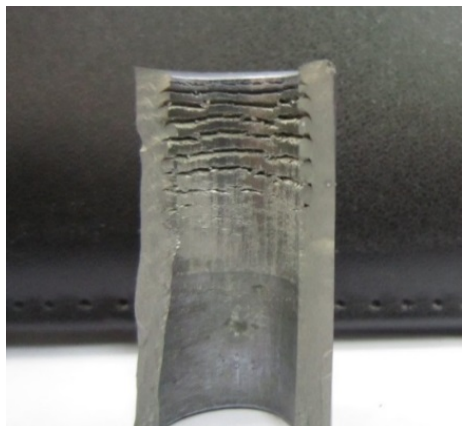

Fig. 36. Image on the vertical section of the backward-extruded ring magnet.

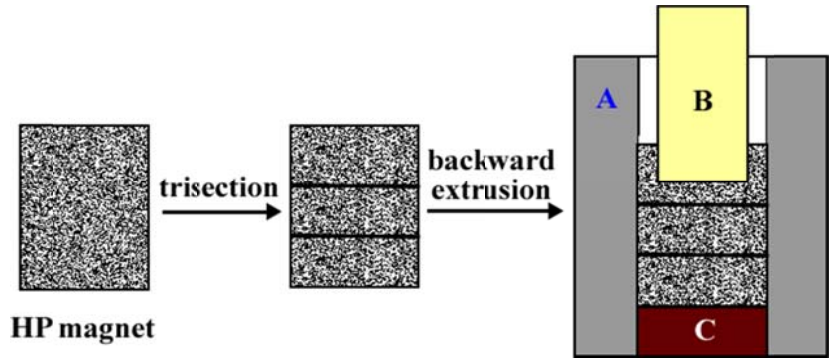

Fig. 37. Schematic illustration on backward extrusion of the stacked HP magnets which were trisected from a mother HP magnet.

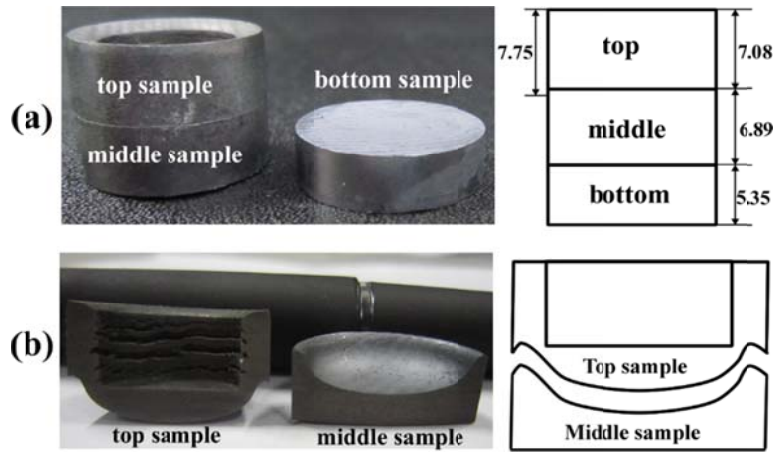

Fig. 38. Photograph and schematic illustration of the bottom (a), middle (b) and top (b) samples after backward extrusion of stacked trisected hot pressing precursor. The height of the ring-shaped part is 7.75 mm.

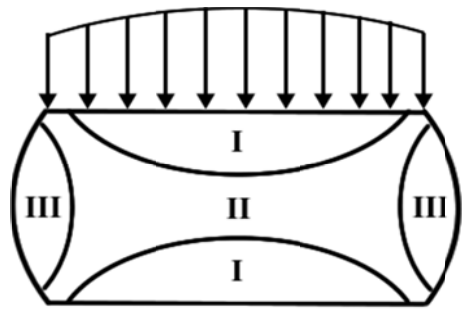

Fig. 39. Illustration on the three zones in cylindrical metals or alloys which were free deformed under uniaxial pressure.

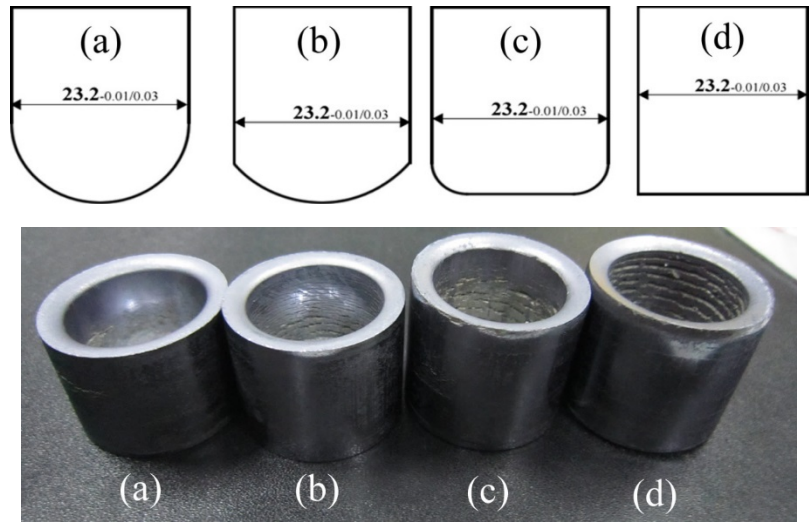

Fig. 40. Images of the different shapes of punches and the ring magnet prepared with them: (a) half-spherical-endpunch; (b) arc-shaped-end punch; (c) beveling cylinder; (d) cylinder.

**Section 6: The physical properties of HD Nd-Fe-B magnets**

6.1 Mechanical properties of HD Nd-Fe-B magnet

Sintered Nd-Fe-B magnet was reported to show the initial worse fracture toughness than metals [92]. Its fracture toughness (KIC) is about 1-2 magnitudes lower than metals [93-94]. This fragile characteristic shorts the life of device and is not good for machinery. HD Nd-Fe-B magnet has the characteristics of both laminated structure consisted of $Nd_2Fe_{14}B$ platelets and anisotropic grain boundary (GB) (Fig.41 and 42). This structure, which is different from that of a traditional sintered Nd-Fe-B magnet, could lead to worse mechanical properties for HD magnet. Jin et al. [95] studied the mechanical properties of HD Nd-Fe-B magnet, revealing a significant anisotropy of mechanical properties (Fig.43). The compression strength, bending strength and fracture toughness along the direction vertical to $Nd_2Fe_{14}B$ platelets are all higher than that parallel to the platelets. The mechanism for this phenomenon can be explained by the laminated structure of $Nd_2Fe_{14}B$ platelets (Fig.44). For HD Nd-Fe-B magnet, the fracture toughness is considered to be the quite important parameter in motor application and machinery. The toughness ($K_{IC}$) along the direction vertical to $Nd_2Fe_{14}B$ platelets is 7.42 $MPa.m^{1/2}$ while that parallel to the platelets is just 1.62 $MPa.m^{1/2}$ (Table 1). In fact, the worse abilities in machinery or application for HD magnets are derived from the worse mechanical properties along the direction parallel to the platelets. Therefore, it is important to raise the toughness parallel to the platelets and/or eliminate the anisotropy of toughness.

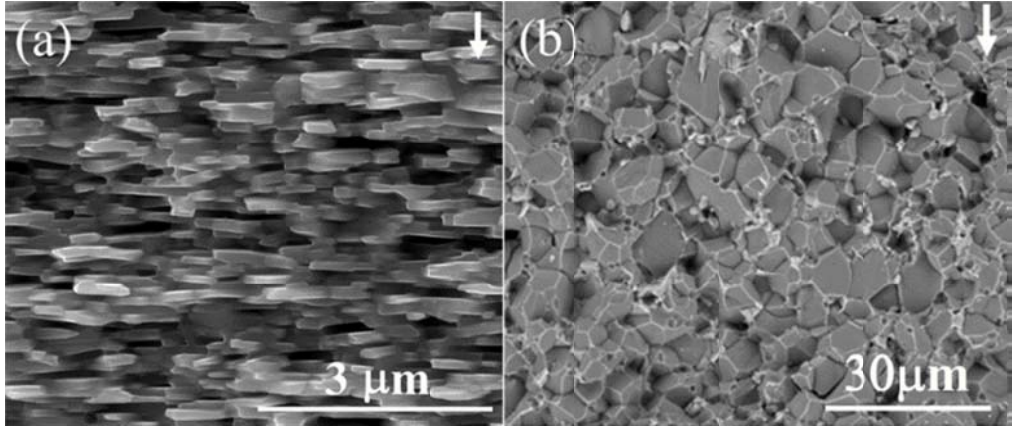

Fig. 41. Microstructure of hot-deformed (a) and sintered (b) Nd-Fe-B magnets

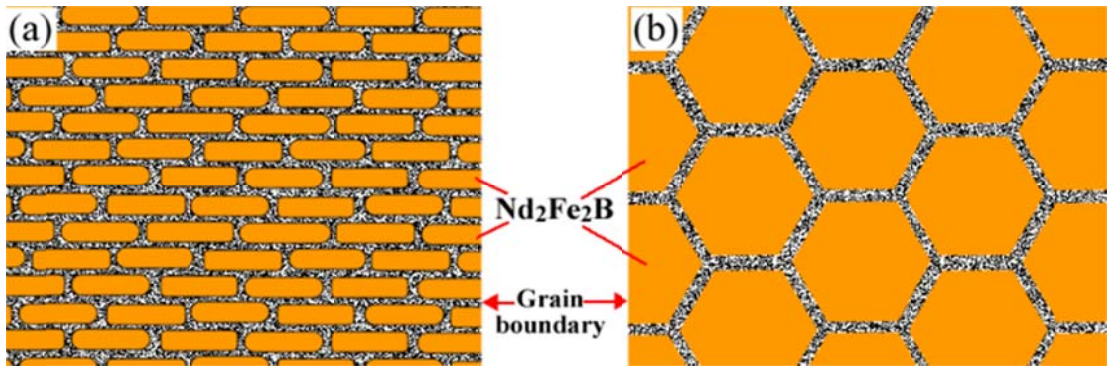

Fig. 42. Schematic illustration on the microstructure of hot-deformed (a) and sintered (b) Nd-Fe-B magnets

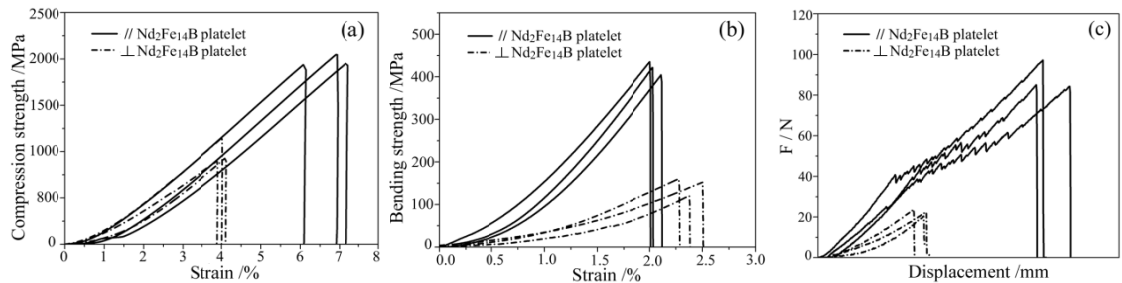

Fig. 43. Compression strength (a), bending strength (b) and fracture toughness (c) of HD Nd-Fe-B magnet along the directions that is parallel (dot-slash line) and vertical (straight line) to $Nd_2Fe_{14}B$ platelets [95].

Table 1. Compression strength ($\sigma_c$), bending strength ($\sigma_b$) and fracture toughness ($K_{IC}$) of HD Nd-Fe-B magnet along the directions parallel (//) and vertical ($\perp$) to $Nd_2Fe_{14}B$ platelets. The deformation ratio of HD Nd-Fe-B magnet is 70%.

| $\sigma_c$(MPa) | | $\sigma_b$(MPa) | | $K_{IC}$(MPa.m$^{1/2}$) | |
|---|---|---|---|---|---|
| // | $\perp$ | // | $\perp$ | // | $\perp$ |
| 1140 | 2060 | 151 | 436 | 1.62 | 7.42 |

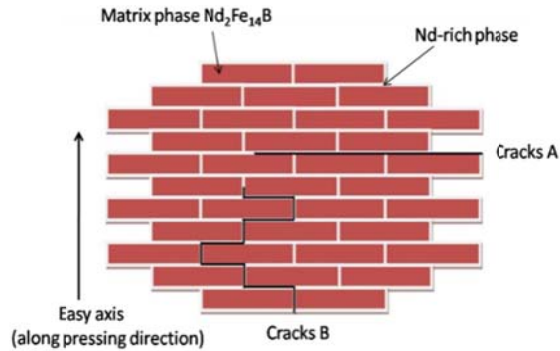

Fig. 44. Sketch map of cracks propagation along grain boundary in HD magnets [95].

Since HD Nd-Fe-B magnets are easily cracked or broken along the GB parallel to $Nd_2Fe_{14}B$ platelets (GB-P), two methods could be carried out to improve the toughness of GB-P: the accumulative precipitation of nano-sized ductile phase and the dispersion of hard metal particles in GB-P (Fig.45). Both the two methods would inhibit the crack propagation or change the path of crack propagation in GB-P, which could raise the mechanical performance along the direction parallel to $Nd_2Fe_{14}B$ platelets for HD Nd-Fe-B magnet. Although such investigation on HD magnet was rarely reported [96-98], similar work has been performed on sintered Nd-Fe-B magnet. Wang *et al.* [99] have reported the enhancement of impact toughness by Dy addition for sintered Nd-Fe-B magnet while the negative effect by Pr addition. Li [100] revealed the positive effect of trace B addition on the bending strength. Other metals and alloys have also been added to improve the mechanical properties of Nd-Fe-B magnet, such as Ti, Co, Ni, $MM_{38}Co_{46.4}Ni_{15.5}$ alloy (MM = Ce-based alloy), Nd, etc. [101-103]. A systematic work about the effect of metal addition (Al, Ga, Cu, Nb, Zr, Ti, V, Mn, etc.) on the impact toughness has been done by Liu *et al.* [104]. Four ideal metals (Al、Ga、Cu、Nb) were proved to be helpful for the increase of impact toughness. The mechanism demonstrates the in-situ precipitation of nano-sized ductile phase in GB after metal addition. The same mechanism was reported by Zeng *et al.* [105], where Nd-(Fe,Co)-Cu and rod-like Nb-Fe-B phases were precipitated and raised the bending strength of Nd-Fe-B magnet after the addition of Cu and Nb, respectively.

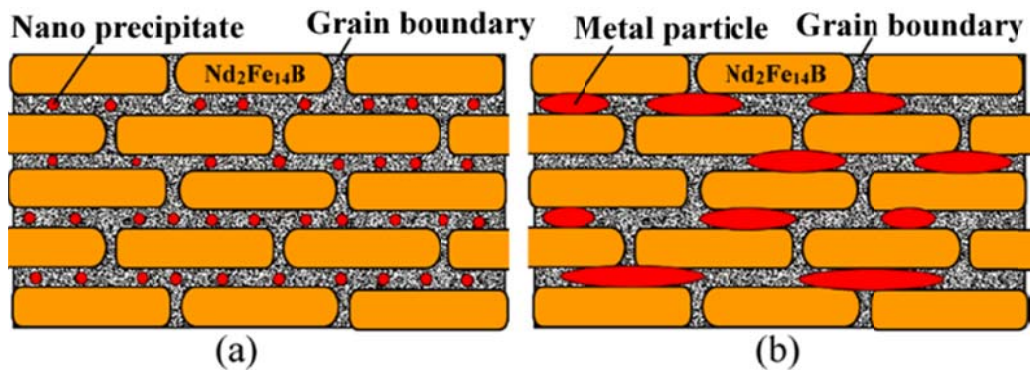

Fig. 45. Illustration on the two method for the enhancement of toughness. (a) Precipitation of nano-sized ductile phase and (b) dispersion of metal particles in the grain boundary parallel to $Nd_2Fe_{14}B$ platelets.

In addition, other preparation parameters have been optimized for better mechanical performance. Yi et al. [17] improved the bending strength of HD magnet by addition of Cu powder. Ju et al. [106] tested the bending strength and toughness for HD magnet with different Nd content, and found that the bending strength perpendicular to $Nd_2Fe_{14}B$ platelets increases with the increment of RE content while that parallel to $Nd_2Fe_{14}B$ platelets exhibits decreasing trend. Lin et al. [16] revealed the coarse grains could deteriorate both the compressing strength and bending strength of HD magnet. This result is consisting with the report that ultrafine grain is good for impact toughness and bending strength [107]. Hu et al. [108-109] studied the mechanical performance over the HD magnet with different deformation rates. It was found that both Vickers hardness and fracture toughness reached the maximum value at the deformation rate of 40%. Hu et al. [110] has also optimized the HD temperature for the Rockwell hardness and fracture toughness. Recently, Zheng et al. [111] found that the cracks of hot-deformed Nd-Fe-B magnets are located in the ribbons boundaries comprise of non-oriented coarse grains. By suppressing the interfacial coarse grains, the mechanical properties can be effectively improved. This can be ascribed to the refined grain size and reduced stress concentration.

6.2 Electric properties of HD Nd-Fe-B magnet

Nd-Fe-B magnet could be assembled in the rotators of brushless motors to get large torque and high efficiency. Nevertheless, Nd-Fe-B magnet is a good conductor and will bring about obvious eddy current loss during rotation in alternating electromagnetic field. The eddy current causes the increase in operating temperature of magnet which reduces the remanence and torque. Therefore, it is favorable to improvement of the magnet stability by suppress the temperature increase in application process.

One commonly-used solution is to raise the electric resistivity of Nd-Fe-B magnet. Since the grain boundary shows much lower resistivity than $Nd_2Fe_{14}B$ phase, it is critical to raise the resistivity of grain boundary. For this sake, Marinescu et al. [112] has introduced fluoride including $NdF_3$, $DyF_3$ and $CaF_2$. The results show that the $CaF_2$ is the most stable addition after hot deformation and brings about the highest resistivity for Pr-Fe-B HD magnet. Furthermore, the coercivity of HD magnet is nearly not decreased. Zheng et al. [113-114] introduced $Al_2O_3$ and $SiO_2$ nanoparticles into Nd-Fe-B HD magnet. Both $Al_2O_3$ and $SiO_2$ nanoparticles could raise the resistivity, but remanence and coercivity decreased significantly. The deterioration of magnetic performance was due to the introduction of oxygen element and the atom diffusion at HD process. After that, Zheng [115, 116] has prepared Nd-Fe-B/$CaF_2$ composite magnet and investigated the effect of $CaF_2$ addition on the properties of HD Nd-Fe-B magnet. Interestingly, $CaF_2$ could transform from discontinuous blocks to continuous laminated structure in the composite magnets when the mixtures were blended in ethanol (Fig. 46). Resistivity of the composite magnet was raised with the increment of $CaF_2$ amount and reached a maximum value of 1280 μΩcm when the content of $CaF_2$ reached 20 vol.%. But the maximum energy product declined to certain level.

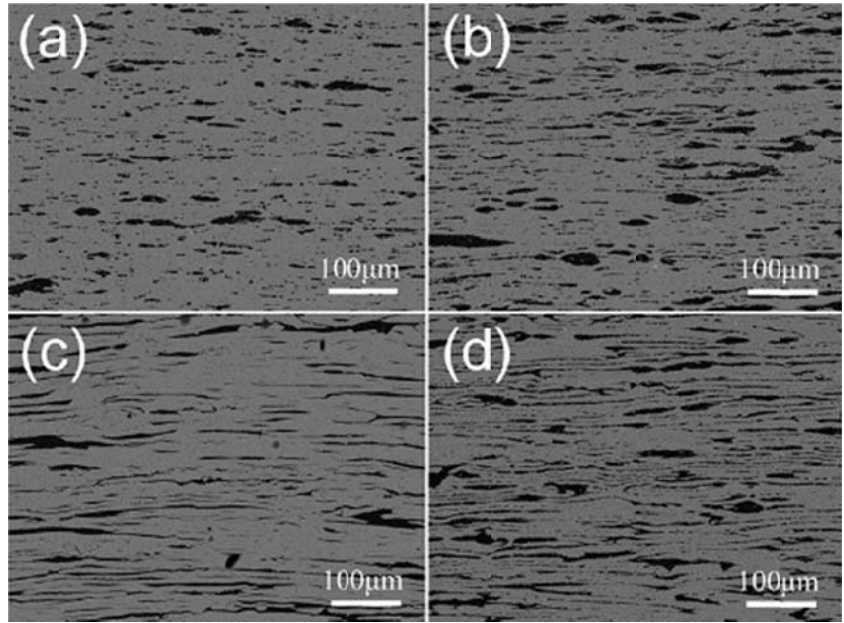

Fig.46. Backscattered electron SEM images of the Nd-Fe-B/CaF$_2$ composite magnet fabricated by (a) simple blending in mortar with 15 vol.% CaF$_2$, (b) simple blending in mortar with 20 vol.% CaF$_2$, (c) wet blending in ethanol and drying with 15 vol.% CaF$_2$, and (d) wet blending in ethanol and drying with 20 vol.% CaF$_2$ [115].

Similarly, Wang *et al.* [117] has developed a facile method in which NdF$_3$ powder was added into Nd-Fe-B magnet in a stacking way (Fig. 47). Electric resistivity increased monotonically with the increase of NdF$_3$ addition, due to the formation of continuous segmented fluoride layers. The electric resistivity increases by nearly eight times when NdF$_3$ content reached 5.3 wt% (Fig.48).

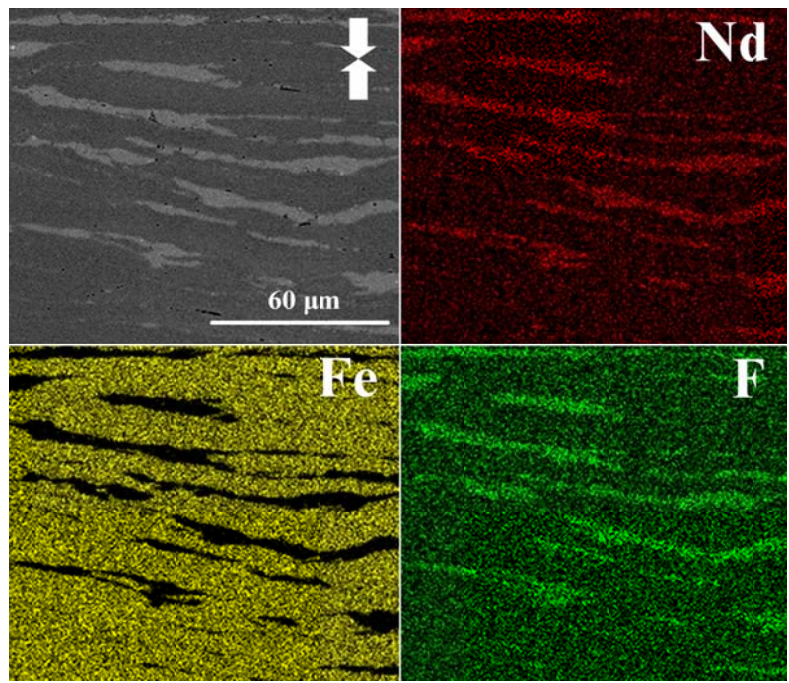

Fig.47. Back scattered electron SEM image for Nd$_2$Fe$_{14}$B/NdF$_3$ (2.12 wt %) hot-deformed magnets with laminated structure and the element concentration maps for Fe, Nd, and F. Arrow is the pressure direction [117].

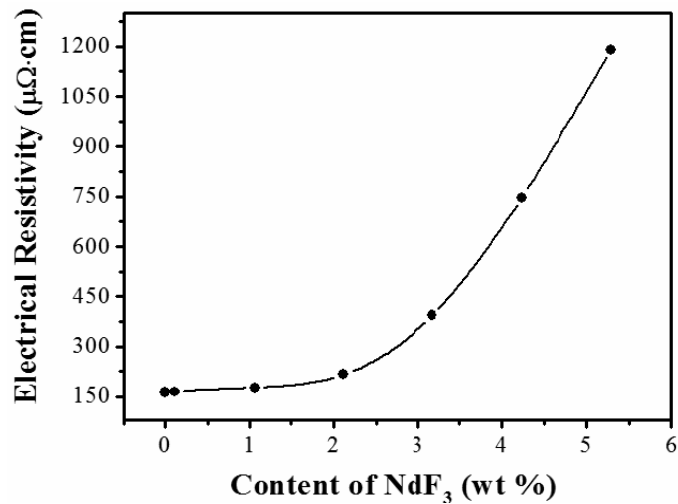

Fig. 48. Dependence of the electrical resistivity of $Nd_2Fe_{14}B$ /$NdF_3$ on the percentage of $NdF_3$ [117].

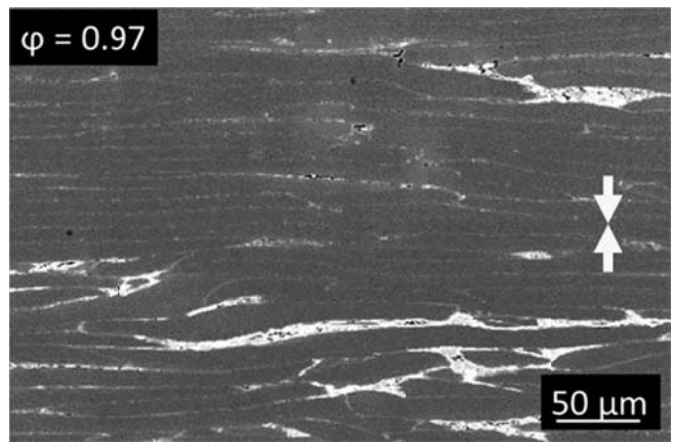

Fig. 49. BSE image on a polished cross section shows the electrically insulating Dy-F rich inclusions (bright area) within the MQU-F flakes (dark areas). The white arrows indicate the pressing direction [118].

Besides, Sawatzki *et al.* [118] also enhanced the electric resistivity of Nd-Fe-B hot-deformed magnets via adding $DyF_3$ powders. The $DyF_3$ additive enriched at the particle boundaries forms insulating Dy-F rich inclusions, which improves the electric resistivity of local regions (Fig. 49).